\documentclass[letterpaper, amsmath,amssymb,preprint]{revtex4-1}
\usepackage{graphicx}
\usepackage{ subfigure, amsmath, amssymb, wasysym, chemarr}
\usepackage{amsmath, amssymb, chemarr}
\usepackage{color}

\begin{document}

\newcommand{\bgamma}{\mbox{\boldmath $\Gamma$}}
\newcommand{\bL}{\mbox{\boldmath $L$}}
\newcommand{\bT}{\mbox{\boldmath $T$}}
\newcommand{\bI}{\mbox{\boldmath $I$}}
\newcommand{\bM}{\mbox{\boldmath $M$}}
\newcommand{\bN}{\mbox{\boldmath $N$}}
\newcommand{\bE}{\mbox{\boldmath $E$}}
\newcommand{\bA}{\mbox{\boldmath $A$}}
\newcommand{\bB}{\mbox{\boldmath $B$}}
\newcommand{\bD}{\mbox{\boldmath $D$}}
\newcommand{\bG}{\mbox{\boldmath $G$}}
\newcommand{\bs}{\mbox{\boldmath $s$}}
\newcommand{\bK}{\mbox{\boldmath $K$}}
\newcommand{\bLambda}{\mbox{\boldmath $\Lambda$}}
\newcommand{\bnu}{\mbox{\boldmath $\nu$}}

\newcommand{\bP}{\mbox{\boldmath $P$}}
\newcommand{\bR}{\mbox{\boldmath $R$}}
\newcommand{\bJ}{\mbox{\boldmath $J$}}
\newcommand{\bp}{\mbox{\boldmath $p$}}
\newcommand{\bx}{\mbox{\boldmath $x$}}
\newcommand{\bU}{\mbox{\boldmath $U$}}
\newcommand{\bV}{\mbox{\boldmath $V$}}
\newcommand{\bZero}{\mbox{\boldmath $0$}}
\newcommand{\bLo}{\mbox{\boldmath $L_0$}}
\newcommand{\bNo}{\mbox{\boldmath $N_0$}}
\newcommand{\bNr}{\mbox{\boldmath $N_R$}}
\newcommand{\bSi}{\mbox{\boldmath $S_i$}}
\newcommand{\bSigma}{\mbox{\boldmath $\Sigma$}}
\newcommand{\bSd}{\mbox{\boldmath $S_d$}}
\newcommand{\bdSi}{\mbox{\boldmath $dS_i$}}
\newcommand{\bdSd}{\mbox{\boldmath $dS_d$}}
\newcommand{\bS}{\mbox{\boldmath $S$}}
\newcommand{\bdS}{\mbox{\boldmath $dS$}}
\newcommand{\bds}{\mbox{\boldmath $ds$}}
\newcommand{\bdx}{\mbox{\boldmath $dx$}}

\newcommand{\bu}{\mbox{\boldmath $u$}}
\newcommand{\bdt}{\mbox{\boldmath $dt$}}
\newcommand{\bdvds}{\frac{\partial \bv}{\partial \bs}}
\newcommand{\bdvdp}{\frac{\partial \bv}{\partial \bp}}
\newcommand{\bdSdt}{\mbox{$\displaystyle \frac{\bdS}{\bdt}$}}
\newcommand{\bv}{\mbox{\boldmath $v$}}
\newcommand{\bw}{\mbox{\boldmath $w$}}

\newcommand{\el}[2]{\varepsilon^{#1}_{#2}}
\newcommand{\uel}[2]{\widetilde{\varepsilon^{#1}_{#2}}}


\title{
Nonlinear  Biochemical Signal Processing via Noise Propagation
}         
\author{Kyung Hyuk Kim$^1$, Hong Qian$^{2,1}$, and Herbert M. Sauro$^1$\\
 {$^{1}$ Department of Bioengineering, University of Washington, William H. Foege Building, 	   Box 355061, 	   Seattle, WA 98195, U.S.A.}\\
{$^{2}$ Department of Applied Mathematics, University of Washington,  Lewis Hall, Box 353925, Seattle, WA 98195, U.S.A.}
}
\date{\today}          
\begin{abstract}
Single-cell studies often show significant phenotypic variability due to the stochastic nature of intra-cellular biochemical reactions.  When the numbers of molecules, e.g., transcription factors and regulatory enzymes, are in low abundance, fluctuations in biochemical activities become significant and such ``noise'' can propagate through regulatory cascades in terms of biochemical reaction networks. Here we develop an intuitive, yet fully quantitative method for analyzing how noise affects cellular phenotypes based on identifying a system's nonlinearities and noise propagations. We observe that such noise can simultaneously enhance sensitivities in one behavioral region while reducing sensitivities in another.  Employing this novel phenomenon we designed three biochemical signal processing modules:  ($a$) A gene regulatory network that acts as a concentration detector with both enhanced amplitude and sensitivity.  ($b$) A non-cooperative positive feedback system, with a graded dose-response in the deterministic case, that serves as a bistable switch due to noise-induced bimodality. ($c$) A noise-induced linear amplifier for gene regulation that requires no feedback. The methods developed in the present work allow one to understand and engineer nonlinear biochemical signal processors based on fluctuation-induced phenotypes.
\end{abstract}

\maketitle

\clearpage

\section{Introduction}
Quantifying biochemical processes at the cellular level is becoming
increasingly central to modern molecular biology \cite{Elowitz2002,
Pedraza2005}. Much of this attention can be attributed to the development
of new methodologies for measuring biochemical events. In particular,
the use of green fluorescent protein  and related fluorophores, high
sensitivity light microscopy and flow cytometry have provided
researchers with greater details of the dynamics of cellular processes
at the level of single cells \cite{Elowitz2002,Pedraza2005,Xie2008a}
with single-molecule sensitivity \cite{Yu2006}.

Quantitative laboratory measurements demand theoretical models
that are able to describe and interpret the data. The traditional approach
to modeling cellular processes has been to use deterministic kinetic equations
with continuous dynamic variables: It is assumed that concentrations of
metabolites, proteins, etc.~vary deterministically in a continuous
manner. This may be a reasonable assumption when one deals with
molecules that occur in relatively large numbers.    In
other cases, the number of molecules can be much lower; for example,
the number of LacI tetrameric repressor proteins in a single
{\it E.~coli} bacterium
has been estimated to be of the order of 10 to 50 molecules
\cite{Oehler1994}. Such small numbers suggest that a continuous-variable
model is no longer an appropriate description of the reality. In
fact, recent experimental measurements have clearly
highlighted the stochastic nature of biochemical processes in
individual cells \cite{Elowitz2002,Choi2008,Qian2012}.

Stochastic reaction processes can be analyzed in terms of probability distributions from an isogenic population or from stochastic trajectories of individual cells.  The former  has been theoretically investigated in terms of the chemical master equation (CME) \cite{Gillespie1992, Mcquarrie1967}.  For general reaction systems, analytically solving this equation is extremely
challenging because the rates of biochemical reactions are often nonlinear functions of concentrations.  In addition, numerical solutions are equally impractical because of the significant increase in the number of states; progress on the computational approach to the CME has been slow \cite{Liang2010}. Alternatively, stochastic trajectories can be computed via Monte Carlo procedures. Methods such as Gillespie algorithm \cite{Gillespie1977} can also be highly intensive computationally for reaction systems involving several
molecular species.   Stochastic dynamics, therefore, are often studied with approximations \cite{Kampen2001, Gomez2007, Scott2007, Munsky2006}.

G\'{o}mez-Uribe et al.~\cite{Gomez2007} provided an approximate method called mass fluctuation kinetics. Recall that in a biological reaction system, the rate of a reaction $v(s)$ is usually a nonlinear function of reactant concentrations, where $s$ is a vector of concentrations.  In terms of cellular biological functions, $s$ is usually referred to as a `signal', and we shall adopt this
terminology.  When the signal is noisy, the expected value of the nonlinear function, $\big\langle v(s) \big\rangle$, is different from the function of the expectation of the concentrations, $v\big(\langle s \rangle\big)$, as illustrated in Fig.~\ref{fig:cur-var-effect} \cite{KimPJ2010}. (Note that for a uni-molecular reaction, $v(s) = ks$ with $k$ being a constant.  Then the mean rate is the same as the rate of the mean concentration: $\langle ks \rangle = k \langle s \rangle$.)  G\'{o}mez-Uribe et al.~\cite{Gomez2007} tried to quantify this difference in terms of both the  the noise strength of  a signal and the degree of reaction nonlinearity with a first-order approximation, assuming the noise strength is sufficiently small. We revisit this nonlinear signal processing of noise and investigate this problem more thoroughly.

Noise in biochemical signals following biological reaction pathways is dependent not only on the `local' reaction rate functions \cite{Fell1992, Kacser1995, Fell1996, Paulsson2004, Sauro2011} but also on the `global' structure of the pathway that the signal `propagates' through \cite{Paulsson2004, Pedraza2005, Hooshangi2005}.  The propagation of noise has been studied mainly by identifying the mathematical analytical structure \cite{Paulsson2004, Tanase2006, Bruggeman2009}. As the system size increases, such a structure quickly becomes intractable.  Here, we  provide a more manageable approach for understanding noise propagation and propose a simple principle we call {\em stochastic focusing compensation}.  {This principle states that a sensitivity increase (stochastic focusing)  \cite{Paulsson2000} in one parameter region is typically achieved by reducing sensitivity in another region due to a geometric constraint.}  To show its wide applicability, we design three different synthetic biochemical networks that exhibit noise-induced phenotypes: (1) noise-enhanced concentration detection, (2) noise-induced bistable ON-OFF switching, and (3) noise-induced linear amplification. These examples illustrate  the proposed method as a powerful tool for designing, analyzing, and understanding noise-induced
phenotypes.


\section{Results}

\subsection{Model systems.}
When molecules undergo chemical reactions particularly at very low concentrations, changes  in amounts are more conveniently described using discrete numbers to represent individual molecules \cite{Kampen2001}.   The random nature of reaction events also dictates the changes being stochastic.   Thus, the time evolution of a system is described by discrete-state continuous-time stochastic processes.  Under appropriate  conditions (spatially homogeneous and statistically independent reaction events), the stochastic processes are  fully described by the CME \cite{Gillespie1992}, or more precisely the Delbr\"{u}ck-Gillespie process \cite{Qian2011}.  The CME describes the time evolution of a probability distribution function, which represents the probability that one finds the number of molecules for each species at a given time.

\subsection{Elasticity.}
Our analysis is based on the concept of elasticity ($\varepsilon$), which quantifies the change in a reaction rate due to a change in a reactant, product or effector concentration.  Mathematically, the sensitivity is  defined as the ratio of the fold change in the rate $v$  to the fold change in  the concentration $s$:
\[
\varepsilon_d = \frac{s}{v}\frac{\partial v}{\partial s} = \frac{\partial \ln v}{ \partial \ln s}.
\]
$\varepsilon_d$ has been widely used in metabolic control analysis \cite{Fell1992, Kacser1995, Fell1996} and here we expand its application to the stochastic regime.

\subsection{Elasticity vs. noise propagation.}
\begin{figure}[t]
\begin{center}
\includegraphics[width=18cm, angle=0]{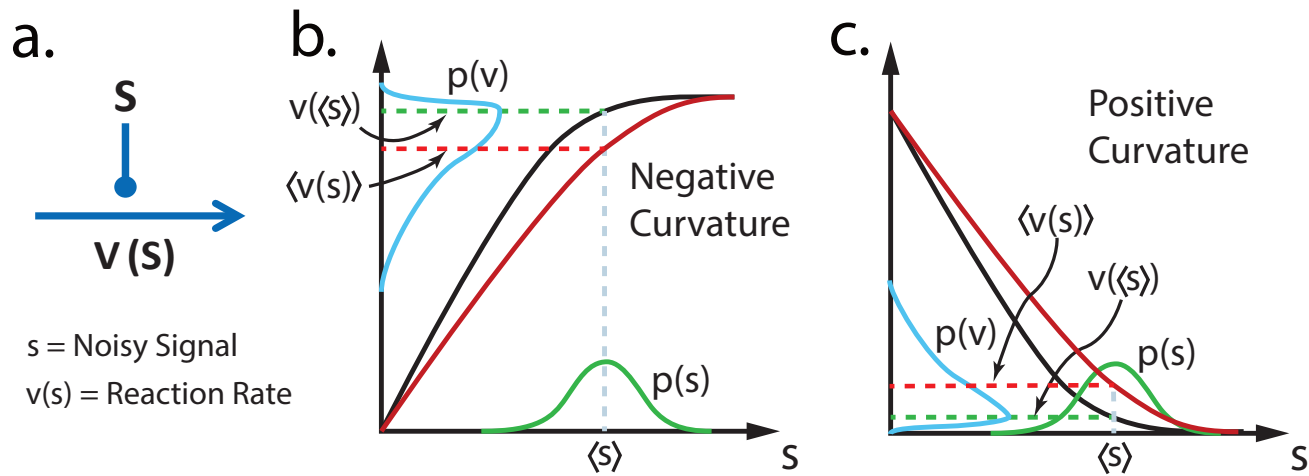}
\end{center}
\caption{  {(a) Signal noise in $s$ is propagated to a reaction step $v(s)$. (b and c) The distribution function of  $s$ is assumed to be symmetric (green). The nonlinear reaction rate (black) results in an asymmetric distribution function of the reaction rate, $v$ (sky blue).  This causes the mean rate (dotted red line) to differ from the deterministic rate (dotted green line).  The curvature of the rate function is negative in (b) and positive in (c). That is, the mean rate (red) becomes smaller or larger than the predicted deterministic rate (black) depending on the sign of the curvature. Mathematically this is a statement of Jensen's inequality.}}
\label{fig:cur-var-effect}
\end{figure}

The mean values of reaction rates can be affected by concentration fluctuations.  This effect was formulated in terms of concentration variances and the curvatures of reaction rate functions \cite{Kampen2001, Gomez2007, KimPJ2010}.  Consider a reaction step shown in Fig.~\ref{fig:cur-var-effect}a. Although the concentration of $S$, denoted by $s$,  is assumed to fluctuate `symmetrically' with respect to its mean value, the reaction rate $v(s)$ fluctuates `asymmetrically' \cite{Ochab2010}.  Let us assume that an increase in $s$ by the amount $\delta s$ causes $v$ to increase by an amount $\delta v$.  A decrease in $s$ by the same amount $\delta s$, however, will not cause $v$ to decrease by the same amount $\delta v$. Instead it is larger than $\delta v$, due to the down-curved shape of the reaction rate function $v(s)$ (Fig.~\ref{fig:cur-var-effect}b).   As the curvature of the reaction rate  function increases, the fluctuations in $v$ become more asymmetric and the mean value of the rate function $\langle v (s) \rangle$ gets smaller, resulting in more deviation from the deterministic prediction $v (\langle s \rangle)$. Here, the angle bracket $\langle \rangle$ denotes ensemble average.   As the distribution function of $s$ gets narrower, the distribution function of $v$ also becomes narrower and the difference between $v(\langle s \rangle)$ and $\langle v(s) \rangle$ smaller.   The first order correction to the deterministic rate was expressed  proportional to both  concentration variance (covariance in general cases) and the curvature of a rate function, and the estimate of the true mean rate $\langle v \rangle$ was expressed  \cite{Kampen2001, Gomez2007}:
\begin{equation}
w(s^*) = v(s^*) + \left.\frac{1}{2}\frac{\partial^2 v}{\partial s^2}\right|_{s=s^*} \Sigma^*,
\label{eqn:MM-deg}
\end{equation}
where $s^*$ is the estimate of the true mean concentration $\langle s \rangle$ and $\Sigma^*$ that of concentration variance  (Methods). The variance in $s$ is dependent on how $s$ is synthesized and degraded as well as other sources of random fluctuations from the connected networks.   {Thus, the interplay between the nonlinearity in the rate function $v$ and the noise in the concentration $s$ causes a change in the mean reaction rate, resulting in changes in elasticities.} Since elasticities together with stoichiometry are the determinants of dynamic behavior, changes in the elasticities due to noise will result in    {changes in system dynamics.} This result allows us to relate deterministic properties of a network to the nonlinear transmission of noise. We will define elasticities that depend on noise propagation as:
\[
\varepsilon_{s} = \frac{\partial \ln \langle v \rangle}{ \partial \ln \langle s \rangle}.
\]
and its first-order approximation as  $\varepsilon_s \simeq \partial \ln w/\partial \ln s^*$.

  {This approximation can be used to derive analytical solutions of mean values and covariances of concentrations, as a next level of the linear noise approximation \cite{Kampen2001},  but needs to be verified on a case by case basis; for example, the level of $X_0$ in Fig.~\ref{fig:ff-diag}a was assumed to be constant but it can be a function of other species concentrations (e.g., Fig.~S2c), which could result in additional noise being propagated into the synthesis rate.}

\subsection{Stochastic Focusing Compensation.}
\begin{figure}[t]
\begin{center}
\includegraphics[width=12cm, angle=0]{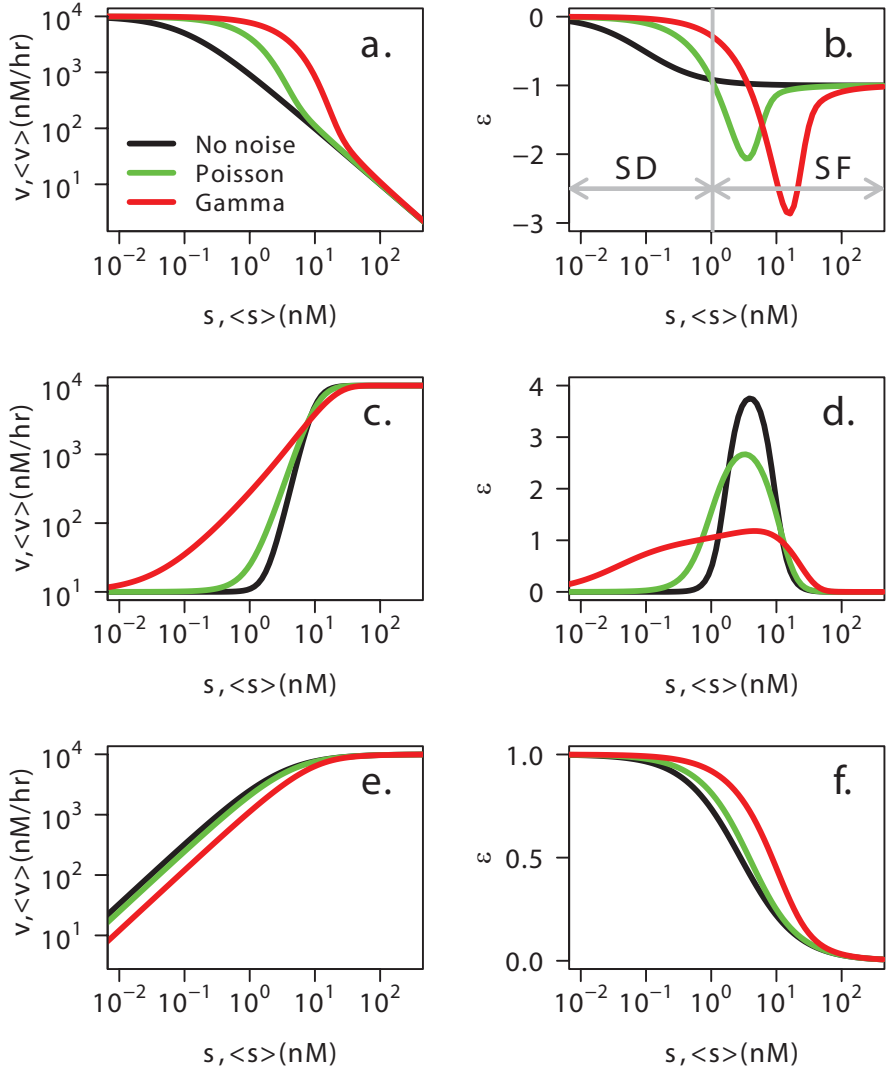}
\end{center}
\caption{Three different types of $v(s)$ show different patterns of compensation. (a,b and c,d) Inhibition and sigmoidal activation lead to stochastic focusing compensation. (e,f) Michaelis-Menten-type activation results in only stochastic focusing (SF) without stochastic defocusing (SD).    {In the subfigure b, SD and SF regions correspond only to the Poisson distribution of $s$ (green line).} The inhibition rate $v(s)$ for (a,b) was given by $10^4/(0.1+s)$, the sigmoidal rate for (c,d) by $10 + 10^4 s^4/(10^4 + x^4)$, and the Michaelis-Menten rate for (e,f) by $10^4 s/(3+s)$. For the distribution  function of $s$, two different distributions were considered: the Poisson distribution, $P(s) = e^{-\lambda}\lambda^s/s!$,  and negative binomial distribution, Eq.~\eqref{eqn:neg} -- a discrete version of the Gamma distribution \cite{Paulsson2000a, Shahrezaei2008}.  The volume of the system was set to $\sim$1~$\mu\mbox{m}^3$ (the cell volume of \emph{E.coli}). 1~nM corresponds to 0.6 molecule per cell, and thus the concentration $s$ in the nM unit was interchanged with the copy number $S$.  The computation was performed in Jarnac~\cite{Sauro2000} and R~\cite{R}.  
}
\label{fig:SDF}
\end{figure}

 Consider the curvature of $v$ in Fig.~\ref{fig:cur-var-effect}c and \ref{fig:SDF}a  which represent  an inhibitory regulation.  The curvature is positive for all $s$ except  when $s\rightarrow 0$ or $\infty$.  Thus, from Eq.~\eqref{eqn:MM-deg} the mean reaction rate $w$ is shown to be larger than the deterministic rate $v$.  When $s\rightarrow \infty$ the curvature approaches zero and when $s \rightarrow 0$ the variance of $s$ vanishes.  Thus,  in these limits $w$ becomes equal to $v$.   Inspection of Fig.~\ref{fig:cur-var-effect}c and \ref{fig:SDF}a shows that the slope of the mean rate ($w$ or $\langle v \rangle$) is smaller than that of $v(s)$ when $s$ is small, implying $\varepsilon_s < \varepsilon_d$.  This elasticity decrease due to noise will be called stochastic de-focusing (SD) (Fig.~\ref{fig:SDF}b). However, as $s$ increases the slope of the mean rate becomes larger than  $v(s)$, implying that $\varepsilon_s > \varepsilon_d$. This sensitivity increase will be called stochastic focusing (SF) (Fig.~\ref{fig:SDF}b). A minor difference from the original definition of SF \cite{Paulsson2000} is discussed in the Appendix.

The previous discussion shows that elasticities that increase due to noise in one parameter region will typically come with elasticity decreases in another region.  Consider the case that $v(s)$ satisfies the Hill equation with Hill coefficient 4 (Fig.~\ref{fig:SDF}c). The curvature of $v(s)$ changes from positive to negative as $s$ increases from zero, and the curvature vanishes as $ s  \rightarrow 0$ or $\infty$.   Thus, defocusing appears in between two focused regions (Fig.~\ref{fig:SDF}d).  In general, focusing does not, however, always come with defocusing. For example, when $v(s)$ is a Michaelis-Menten-type rate function, only focusing appears as shown in Fig.~\ref{fig:SDF}e and f.

When $s$ satisfies the Poisson distribution, focusing and defocusing can occur strongly.  With a different type of distribution such as the Gamma distribution, these can appear even more strongly (Fig.~\ref{fig:SDF}). The Gamma distribution has been frequently found in the case when translation events occur in a burst manner \cite{Taniguchi2010}.  One of the important differences between the Poisson and Gamma distributions is that  in the Gamma distribution one can control the noise (coefficient of variation; CV) and mean levels independently, while one cannot do so in the Poisson distribution \cite{Kim2012}.   {Furthermore, the Poisson distribution is the asymptotic case of the Gamma distribution as the burst size in the translation events reduces to zero.  These two facts  imply that the noise strength (CV) increases from that of the Poisson case  by turning on the bursting events, resulting in a larger deviation from the deterministic case based on Eq.~\eqref{eqn:MM-deg}.}  This is  the reason that stronger focusing and defocusing were observed for the case of the Gamma distribution as shown in Fig.~\ref{fig:SDF}.

In this section, we have shown that higher elasticities in one region of parameters can be achieved by reducing elasticities in another region, and have explained this \emph{stochastic focusing compensation} by using the curvature-variance correction to a mean reaction rate (Eq.~\eqref{eqn:MM-deg}). The compensation was intuitively understood as a geometric effect of the reaction rate. In the following sections we will show how this effect can be used to modify the behavior in gene regulatory networks.

\subsection{Application 1: Noise-improved Concentration Detector.}
\begin{figure}[t]
\begin{center}
\includegraphics[width=15cm, angle=0]{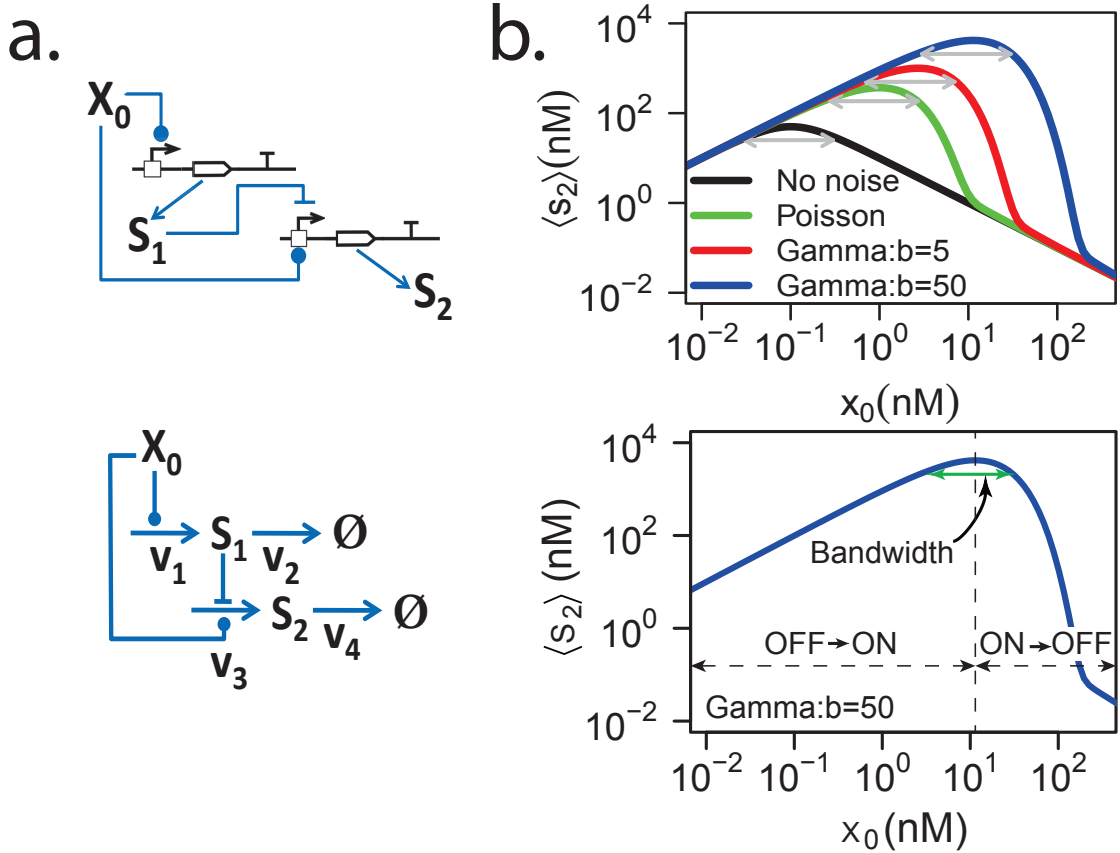}
 \end{center}
\caption{An incoherent feed-forward network is considered. (a) A transcription factor $X_0$ up-regulates the expression of both $S_1$ and $S_2$, and $S_1$ down-regulates the expression of $S_2$. The concentration of $X_0$ is considered fixed (boundary species). Reaction rates: $v_1 = p_1 x_0$, $v_2 = p_2 s_1$, $v_3 = \frac{\alpha p_3/p_4}{s_1 + (1+p_3 x_0)/p_4 x_0}$, and $v_4 = p_5 s_2$. (b) Stochastic focusing compensation was used to improve the detection amplitude and ON$\rightarrow$OFF sensitivity. With larger noise (Gamma: $b=5$, $b=50$), the detection amplitude was further enhanced. Parameters for (b): $p_1=100~\mbox{hr}^{-1}$, $p_2=100~\mbox{hr}^{-1}$, $p_3=0.01~\mbox{nM}^{-1}$, $p_4=100~\mbox{nM}^{-2}$, $p_5=10~\mbox{hr}^{-1}$, and $\alpha = 10^5~\mbox{nM/hr}$. The computation was performed in Jarnac~\cite{Sauro2000} and R~\cite{R}. The gene regulatory network was represented by using the Synthetic Biology Open Language Visual \cite{sbolv}.
}
  \label{fig:ff-diag}
\end{figure}

Let us consider an incoherent feed-forward gene regulatory network that functions as a concentration detector \cite{Entus2007, Kaplan2008}. We can use noise to enhance the detection amplitude and sensitivity of this system.  Consider a reaction process where a protein ($S_2$ in Fig.~\ref{fig:ff-diag}a) is regulated by two different pathways: either via direct transcriptional activation by $X_0$ or indirect inhibition by $S_1$. When the concentration of $X_0$ is zero, $S_2$ is not expressed.  As $x_0$ increases, $s_2$ increases together (this region of $x_0$ will be named OFF$\rightarrow$ON as shown in Fig.~\ref{fig:ff-diag}b).   When $x_0$ becomes larger than a thresh-hold point, it begins, however, to decrease and is eventually dominated by $S_2$'s inhibition (this region of $x_0$ will be named ON$\rightarrow$OFF as shown in Fig.~\ref{fig:ff-diag}b). Thus, one can detect a specific range of the concentration of $X_0$ by monitoring the concentration of $S_2$.  The plot of $x_0$ vs. $s_2$ (Fig.~\ref{fig:ff-diag}b) will be used to characterize the concentration detection by using the slope as sensitivity, its width at the half maximum as bandwidth, and the peak value of $s_2$ as amplitude.

We will apply stochastic focusing compensation to enhance the detection amplitude and sensitivity.    {In this system, noise in the concentration levels of $S_1$ and $S_2$  is generated from a series of stochastic reaction events of  synthesis and degradation of  $S_1$ and $S_2$.  Since $S_1$ regulates the synthesis of $S_2$, noise in the concentration level of $S_1$ introduces additional noise into $S_2$, i.e., the noise propagates. Based on the fact that  the regulation is inhibitory}, we can conclude that both strong SF and SD can appear in the ON$\rightarrow$OFF region (refer to Fig.~\ref{fig:SDF}a,b). For simplicity, here we assumed that the concentration of $X_0$ does not fluctuate stochastically and thus the activation by $X_0$ does not have any noise effect. Therefore, we focus on the inhibition pathway and aim to increase the detection sensitivity in the ON$\rightarrow$OFF region.

For this aim, strong SF is required to appear in the ON$\rightarrow$OFF region, while minimizing the appearance of SD in the same region. In the hyperbolic inhibition case (Fig.~\ref{fig:SDF}a), strong SF was achieved for the value of $K_M \lesssim 0.1$~nM  and for that of $S_1$ between 1 and 10~nM (for the Poisson case in Fig.~\ref{fig:SDF}a,b) \cite{SI}.  Thus, we tuned the values of the terms corresponding to $K_M$, $(1+p_3 x_0)/p_4 x_0$, to less than 0.1~nM and the mean value of $s_1$, $p_1 x_0/p_2$, between 1 and 10~nM.  These two conditions provided the possible range of $x_0$: $\mbox{MAX}\left( \frac{1}{0.1 p_4-p_3}, \frac{p_2}{p_1}  \right) < x_0 <  10\left(\frac{p_2}{p_1}\right)$, where the function MAX chooses the larger value of its parameter arguments.

The above $x_0$ region was placed to the right side of the detection peak but not too far away from the peak to maximize the appearance of strong SF and to minimize the effect of strong SD in the region. Then, concentration detection was enhanced 5 times in the ON$\rightarrow$OFF sensitivity and 7 times in the amplitude (Fig.~\ref{fig:ff-diag}b) due to the strong SF \cite{SI}. However, the bandwidth stayed the same (arrows in Fig.~\ref{fig:ff-diag}b), which is related to the stochastic focusing compensation; SD appears as $x_0$ decreases away from the region of SF, resulting in reduced detection sensitivity and eventually preventing the detection bandwidth from decreasing.

To enhance the detection amplitude and sensitivity further, we used the Gamma distribution to allow larger noise levels in $s_1$. As the burst size $b$ -- number of translation events from a single mRNA during its average lifetime -- increases, the amplitude of the detection is significantly enhanced: 20 times for $b=5$, 83 times for $b=50$ compared to the deterministic case. However, when we allow $x_0$ to fluctuate stochastically, its signal noise turned out to slightly decrease the amplitude due to the positive curvature contribution in the activation by $X_0$ as well as the noise correlation between $x_0$ and $s_1$ \cite{SI}. This further analysis shows that our curvature-variance-based analysis can successfully predict the change in the system properties.

\subsection{Application 2: Noise-induced Bistable ON-OFF switch.}
Bistability is a common feature of many biological systems  \cite{Ozbudak2004, Brandman2005}.  In the stochastic framework, bistability can sometimes be observed  as a bimodal distribution because the system can jump from one stable state to the other.  In this section we will show how a non-bistable system can be made bistable through stochastic focusing  compensation. This example shows the predictive power of our method in system design and furthermore provides a novel mechanism of noise-induced bistability.

\begin{figure}[t]
\begin{center}
\includegraphics[width=18cm, angle =0]{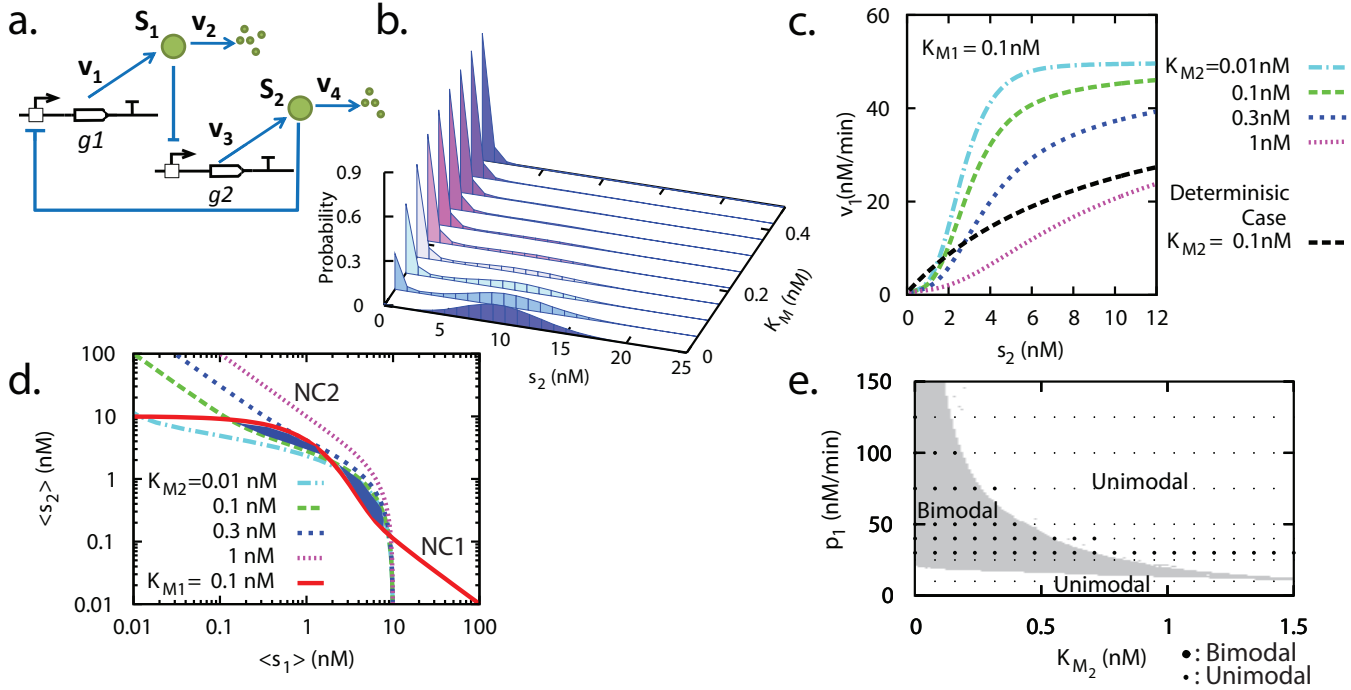}
\end{center}
\caption{ (a) Two genes mutually inhibit each other non-cooperatively. (b) Bimodal distributions in $S_1$ (not shown here) and $S_2$ are observed in the presence of noise. (c) Stochastic focusing compensation caused ultra-sensitive positive feedback. (d) Two nullclines corresponding to $\langle v_1 \rangle  = p_2 \langle s_1\rangle$ and $\langle v_3 \rangle =  p_4 \langle s_2 \rangle$ are shown. The nullclines were computed using the Poisson distributions for $s_1$ and $s_2$ as a first level of approximation. The doublet of stochastic focusing and de-focusing caused bimodality. (e) Under the approximation, a phase plot of distribution modality was obtained (gray and white areas). It matched well with exact stochastic simulation results (dots) based on Gillespie's algorithm \cite{Gillespie1977}. $v_1 = p_1/(1+s_2/K_{M_2})$, $v_2 = p_2 s_1$, $v_3 = p_3/(1+s_1/K_{M_1})$, and $v_4 = p_4 s_2$ with $p_2=p_4=5$~hr$^{-1}$, $p_3 = 50$~nM/hr, and $K_{M_1} = 0.1$~nM. In (b,c,d), $p_1 = 50$~nM/hr.
}
\label{fig:bimodal}
\end{figure}

We consider the model of mutual inhibition between two genes (\emph{g1} and \emph{g2} in Fig.~\ref{fig:bimodal}a), which regulate each other non-cooperatively (Hill coefficient = 1).  Under these conditions it is impossible for the system to show bistability  with first-order degradation in the deterministic case \cite{Sauro2009}.  Now we consider including noise in the system and show noise-induced bistability. The dissociation constant of $S_2$, $K_{M_2}$, was varied while that of $S_1$, $K_{M1}$, was fixed at  $0.1$~nM. The reason that this $K_{M1}$ value was chosen is  to achieve strong SF and SD in the synthesis rate of $S_2$, $v_3$. For $K_{M_2} \simeq $~0.1~nM, another strong SF and SD in $v_1$ appears. This doublet of SF-SD can lead to ultra-sensitive positive feedback (Fig.~\ref{fig:bimodal}c), leading to bimodality (Fig.~\ref{fig:bimodal}b).  As $K_{M_2}$ increases, the SF-SD in $v_1$ becomes weaker and the bimodal distribution becomes uni-modal.

To understand how the doublet of SF-SD causes bimodality, the nullclines of $d\langle s_1 \rangle/dt$ and $d\langle s_2 \rangle /dt$ were plotted in Fig.~\ref{fig:bimodal}d.    {The nullclines were computed using the Poisson distributions for $s_1$ and $s_2$ as a first level of approximation.}  In Fig.~\ref{fig:bimodal}d, the solid line (NC1) corresponds to $\langle v_3 \rangle = p_4 \langle s_2 \rangle$ with $K_{M_1} = 0.1$~nM  and  was significantly distorted due to the strong SF-SD (cf. Fig.~\ref{fig:SDF}a). With $K_{M_2} = 0.1$~nM, the other nullcline (NC2) corresponding to $\langle v_1 \rangle = p_2 \langle s_1 \rangle$ was also significantly distorted. These two distorted curves, NC1 and NC2, crossed each other by making a shape of $\infty$. As $K_{M_2}$ increases, NC2 became straightened out due to weaker SF-SD and  shifted upward. Thus the bimodality disappeared.

The above analysis is limited to unimodal distributions and thus significant care may need to be taken when the distribution deviates from unimodality and enters into bimodality.   {Therefore, we checked the numerical accuracy of our analysis on this system.} The phase diagram for bimodality and unimodality  were plotted for different values of $p_1$ and $K_{M_2}$. The predicted phase diagram (gray and white areas in Fig.~\ref{fig:bimodal}e) matched well with the results based on exact stochastic simulations (dots in Fig.~\ref{fig:bimodal}e).    {This shows that the noise-induced bimodality appears due to the interplay between the system nonlinearity and stochasticity that are described by the CME description and } that our method based on stochastic focusing compensation can be used for predicting bimodality in a rational way.

\subsection{Application 3: Noise-induced Linear Amplifier.}
A linear amplifier is a device that transfers an input signal to an output without distorting the signal shape, resulting in reliable information transfer. Such devices have been observed in biological systems, for example, in signal transduction pathways in yeast cells (\emph{Saccharomyces cerevisiae})  \cite{Yu2008} and mammalian cells \cite{Sturm2010}. This linear amplification has been attributed to negative feedback mechanisms  \cite{Sauro2007, Sturm2010}.  Here we present a novel way to achieve linear amplification by exploiting noise without resorting to any feedback.

The SF-SD makes a sigmoidal curve less steep as illustrated in Fig.~\ref{fig:SDF}c. The contribution becomes larger as the strength of noise increases and the SF-SD can appear more strongly (Eq.~\eqref{eqn:MM-deg}), i.e., a transfer curve can become less sigmoidal with the increase of noise strength. Based on this prediction, we investigated how gene expression noise can be leveraged to build a linear amplifier without using any feedback.

\begin{figure*}[t]
\includegraphics[width=18cm, angle =0]{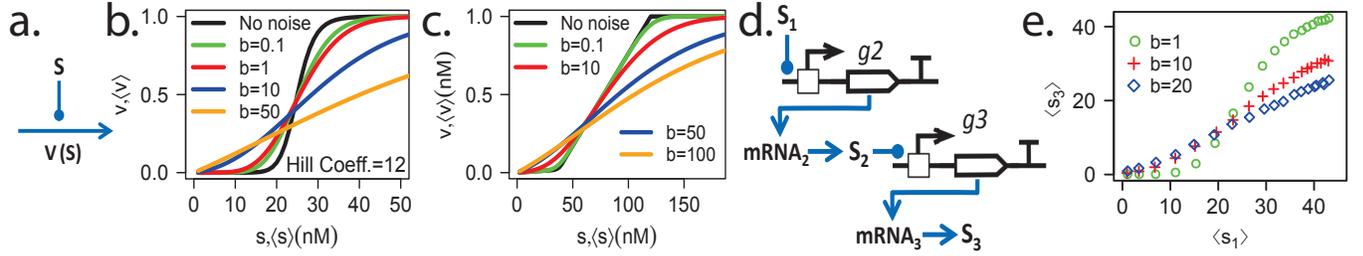}
\caption{ (a) $S$ activates the downstream reaction $v(s)$. (b and C) As the strength of noise increases,  $\langle v(s) \rangle$ becomes more linearized; (b) $v(s)$ is a sigmoidal function defined by $v_{max} s^{12}/(25^{12} + s^{12})$ with $v_{max} = 1$~nM/h  and $b$ is the burst size by considering $S$ as a transcription factor that follows the Gamma distribution. (c) $v(s)$ is a piece-wise linear function (for its detailed functional form, refer to \cite{SI}). (d) A gene activation cascade was considered including transcription processes. The two transcription activations are the same except they are regulated by different transcription factors (refer to \cite{SI} for the model description) (e) The response between $S_1$ and $S_3$ became linearized as the burst size ($b$) increases.
}
\label{fig:linear}
\end{figure*}

We considered   {two different activation reaction systems} (Fig.~\ref{fig:linear}a): a Hill equation with Hill coefficient = 12 (Fig.~\ref{fig:linear}b) and a piece-wise linear function (Fig.~\ref{fig:linear}c). An activator $S$ is assumed to satisfy the negative binomial distribution -- a discrete version of the Gamma distribution:
\begin{equation}
P_s = \frac{\Gamma(a+s)}{\Gamma(s+1)\Gamma(a)}\left( \frac{b}{1+b}\right)^s\left( 1- \frac{b}{1+b}\right)^a
\label{eqn:neg}
\end{equation}
where $a$ is a burst frequency (number of transcription events during protein lifetime), ranging between 0.1 and 11 in \emph{E. coli} with a YFP-fusion tag \cite{Taniguchi2010}, and $b$ a burst size, ranging between 1 and 900 \cite{Taniguchi2010}. Here the mean of the distribution is equal to $ab$ and the strength of noise (standard deviation divided by mean)  is $\sqrt{(1+b)/ab} \simeq 1/\sqrt{a}$ for $b \gg 1$. We varied the strength of noise by increasing $b$ (more precisely for a fixed mean value, increasing $b$ corresponds to decreasing $a$, resulting in an increase in the noise strength) and observed the change in $\langle v(s) \rangle$. It turned out that $\langle v(s) \rangle$ became dramatically linearized as $b$ increased (Fig.~\ref{fig:linear}b; cf. Fig.~3 in \cite{Thattai2002}). More surprisingly, this linearized response was observed not only for the Hill function-type activation with Hill coefficients ranging from 1 to 20 \cite{SI}, but also holds for various types of activation patterns \cite{SI} including piece-wise linear functions (Fig.~\ref{fig:linear}c).

To confirm this noise-induced linear amplification, stochastic simulations were performed for a three-gene cascade (Fig.~\ref{fig:linear}d) by using the Gillespie stochastic simulation algorithm,  with transcription processes included. We  achieved a linear response for the transfer curve for $s_1$ vs. $s_3$ (Fig.~\ref{fig:linear}e).  This result implies that gene expression noise generated upstream can linearize downstream nonlinear transcriptional responses.  This leads to the non-intuitive result that signals upstream can be transferred reliably without distortion as a result of noise.  This  noise-induced linearization may be observed in protein signaling pathways as well if the protein concentration distributions are wide \cite{Birtwistle2012, Thattai2002}.  This could serve as an alternative mechanism for dose-response alignment phenomenon \cite{Yu2008, Sturm2010} without strongly resorting to negative feedback.

\section{Discussion}
We analyzed how noise propagation, in the steady state of stochastic biochemical reaction systems \cite{Paulsson2004, Pedraza2005, Hooshangi2005}, affects the population average of reaction rates and thus their elasticities (sensitivities).  The change in the elasticity was explained in terms of a  curvature-variance contribution, which led us to discover stochastic focusing compensation: A (stochastic) elasticity increase in one parameter region due to noise can lead to an elasticity decrease in another region.

This stochastic focusing compensation was used to enhance the functional properties of a concentration detector having an incoherent feedforward network structure. We designed strong stochastic focusing to a desired detection region and improved the detection amplitude and sensitivity. The detection bandwidth was, however, not enhanced due to stochastic defocusing
In addition, we applied the same idea of stochastic focusing compensation in designing a noise-induced bistable ON-OFF switch
that has a non-cooperative positive feedback loop (thus showing a single stable state in the absence of the noise). The underlying mechanisms of the emergent bimodality were related to the doublet of stochastic focusing-defocusing that appear in two individual inhibitory interactions, with the two as a whole comprising a positive feedback loop. Lastly, the stochastic focusing compensation principle was also applied to the design of a noise-induced linear amplifier. Surprisingly, the biologically-relevant distribution of proteins (Gamma distribution) were shown to have important properties that linearize a wide range of sigmoidal responses.

In our linear amplification study, it seems paradoxical that greater noise enables more reliable information transfer. However, the fact that the linearization becomes saturated as the noise strength (CV) increases (refer to \cite{SI}) implies that there will be an optimal level of noise satisfying optimal signal transfer much like stochastic resonance \cite{Gammaitoni1998} but with a completely different mechanism.

Experimental evidence of the Gamma distribution of intracellular protein copy numbers \cite{Paulsson2000a, Friedman2006, Shahrezaei2008, Taniguchi2010, Birtwistle2012a} implies that when measuring the Hill coefficient of transcriptional regulation, significant care may need to be taken  to include this stochastic focusing compensation. Furthermore, in signal transduction networks, cell-to-cell variability in receptor occupancies and intra-cellular enzyme copy numbers \cite{Birtwistle2012, Birtwistle2012a} could lead to lowering the observed Hill coefficients of downstream regulation and contributing to dose-response alignment. It would be important to investigate the degree of individual contributions of this noise effect (SF-SD) and negative feedback loops \cite{Yu2008}.


\section{Conclusions}
Noise propagation can have significant effects on cellular phenotypes \cite{Arkin1998, Maamar2007, Wang2011}.  It is important to be able to explain propagation mechanisms and to use them for designing or adjusting phenotypes \cite{Kim2012}. We provide a manageable, yet fully quantitative approach for studying noise propagation and its effect, in terms of the degree of a system's nonlinearity and the strength of noise. We show that an increased sensitivity in one parameter region often comes with a decreased sensitivity in another parameter region. We applied this sensitivity compensation characteristics in designing three gene regulatory networks: a concentration detector, a noise-induced ON-OFF switch, and a noise-induced linear amplifier. These examples illustrate how our sensitivity-based approach can be used to design, analyze, and optimize the functionalities of stochastic reaction networks by exploiting noise.

\section*{acknowledgments}
We acknowledge the support from the National Science Foundation (Theoretical Biology 0827592 and Molecular and Cellular Biosciences 1158573; for Preliminary studies, FIBR 0527023).

\appendix
\section{Model Systems}
Random chemical reactions are often modeled by stochastic processes governed by the chemical master equation (CME) \cite{Gillespie1992}, under the assumption that the reactions occur uniformly in position space and independently.   This equation describes the time evolution of a probability distribution function of reactant numbers for all species at a given time.

We consider $m$ species of molecules involved in $n$ reactions:
\begin{eqnarray*}
n_1^l S_1 + \cdots + n_m^l S_m \xrightarrow{V_l} m_1^l S_1 + \cdots + m_m^l S_m,
\end{eqnarray*}
where the molecule numbers are denoted by $\{S_i\}$ with $i=1, \cdots, m$ and the rate of reaction by $V_l$ with $l = 1, \cdots, n$.  We also assume that  the rate of reactions $\bf{V}$, denoting  \{$V_1, \cdots, V_n$\},  can be controlled by changing  a set of system parameters $\bp$.  The parameters are non-fluctuating (bold symbols represent matrices and vectors).  The number change in a species $i$ by a single event of the above reaction $l$ is described by a reduced stoichiometry matrix: $N_{R_{il}} \equiv m_i^l-n_i^l$.    The molecule numbers evolve stochastically and their evolution can be described by the CME:
\begin{eqnarray}
\frac{\partial P\big(\bS,t\big)}{\partial t}& =& \sum_{j=1}^n \Big[ P\big(\{S_i-N_{R_{ij}}\},t\big)V_j\big(\{S_i-N_{R_{ij}}, \bp\}\big) \nonumber\\
&& - P(\bS,t)V_j(\bS,\bp) \Big],
\label{app:master}
\end{eqnarray}
where $\bS$ denotes $\{S_i\}$ with $i=1, \cdots, m$, and $P(\bS, t)$ is  a probability distribution function of reactant numbers for all species at a given time.

\section{Time evolution of concentration mean values and covariances}
We switch the representation of states from numbers $\{ \bS \}$ to concentrations of  molecules $\{\bs \equiv \bS/\Omega \}$ with $\Omega$ a system volume, since this concentration representation has a direct correspondence to deterministic macroscopic kinetics.

The mean concentration of a species $i$  is given as
\[
\langle s_i \rangle_t \equiv \left[\prod_{j=1}^m \int_{s_j=0}^ \infty ds_j\right] s_i P(\bs,t),
\]
where $P(\bs,t)$ is the probability distribution function of molecule concentrations, $\bs= \{s_1, s_2, \cdots, s_m\}$, at a given time $t$.  We will use the angle bracket $\langle \cdot \rangle_t$ to represent the ensemble average at a given time $t$ from now on. The evolution of the mean concentration of a species $i$ is governed by the following equation \cite{SI}:
\begin{equation}
\frac{d\langle \bs \rangle_t }{dt} = \bN_{R}  \langle \bv(\bs,\bp) \rangle_t,
\label{app:mean}
\end{equation}
 where $\bN_R$ is a $m_0 \times n$ reduced stoichiometry matrix \cite{Reder1988, Sauro2004}.  $m_0$ is the number of linearly independent rows of a stoichiometry matrix. $\bv$ represents rate functions: $\bv\equiv \{ v_1, \cdots, v_n\}$ with $v_i \equiv V_i/\Omega$.   We note that $v_i$ is a function of $\{s_j\}$ with $j=1, \cdots, m_0$ and $\bp$.

A concentration covariance between two species ($i$ and $j$) is defined as
\[
\Sigma_{ij}^t \equiv \Big\langle (s_i -\langle s_i \rangle_t )(s_j -\langle s_j \rangle_t )\Big\rangle_t.
\]
The angle brackets again represents ensemble averages at a given time $t$.
 The evolution of the covariance matrix is  described by the following equation \cite{SI}:
\begin{equation}
\frac{d \bSigma_t  }{dt} = \Big\langle (\bN_R \bv)  (\bs-\langle \bs \rangle_t) + (\bs-\langle \bs \rangle_t)^T  (\bN_R \bv)^T + \frac{\bD}{\Omega} \Big \rangle_t,
\label{app:variance}
\end{equation}
where the diffusion coefficient matrix $\bD$ is defined by $ \bN_{R} \bLambda \bN_{R}^T $ with a diagonal matrix $\Lambda_{ij} \equiv v_i\delta_{ij}$.

It  is difficult to solve equations~(\ref{app:mean}) and (\ref{app:variance}) unless the rate function $\bv(\bs,\bp)$ is linear with $\bs$.  Under the assumption that the third and higher order moments of $s-\langle s \rangle_t$ are negligible, the time evolution  of concentration mean values and covariances can be described by the following equations \cite{Gomez2007} \cite{SI}:
\begin{eqnarray}
\frac{d\langle \bs \rangle_t }{dt} &=& \bN_R  \Bigg(\bv( \langle \bs \rangle_t, \bp  ) + \sum_{ij}\frac{1}{2}\frac{\partial^2 \bv(\langle \bs \rangle_t, \bp ) }{\partial \langle s_i \rangle_t \partial \langle s_j \rangle_t}  \Sigma_{ij}^t\Bigg),
\label{mean1}
\end{eqnarray}
\begin{eqnarray}
\frac{d\bSigma_t}{dt} &=&   \bJ \bSigma_t+  \bSigma^T_t  \bJ^T + \frac{ \bD(\langle \bs \rangle_t ) }{\Omega},  \label{variance1}
\end{eqnarray}
where $J_{ij}$ is an element of the Jacobian matrix defined as
\[
J_{ij}\equiv \sum_{k=1}^n N_{R_{ik}}  \frac{\partial v_k(\langle \bs \rangle_t, \bp )}{\partial \langle s_j \rangle_t}.
\]
The validity of the above equations  should be checked on the basis of  each different model, since noise in adjacent systems can be propagated into the rate function \cite{Paulsson2004, Pedraza2005, Hooshangi2005}.  We note that Eq.~\eqref{variance1} is different from the results of G\'{o}mez-Uribe, et al.~\cite{Gomez2007}: We have neglected all the terms of the order of $1/\Omega^2$ that have been kept in Eq.~11 of G\'{o}mez-Uribe, et al. \cite{Gomez2007}.    This is consistent within the approximation of  truncation of third and higher order moments.

\section{Mean reaction rates and covariance matrices at the stationary state}
From Eq.~\eqref{mean1}, the mean rate of reaction can be approximated as:
\begin{equation}
\bw \equiv \bv(\langle \bs\rangle, \bp) + \sum_{ij} \left.\frac{1}{2}
   \frac{\partial^2 \bv}{\partial s_i\partial s_j}\right|_{\bs=\langle \bs\rangle}
\Sigma_{ij}^t.
\label{eqn:vtilde}
\end{equation}
At the stationary state, Eqs.~\eqref{mean1} and \eqref{variance1} becomes:
\[
\bN_R \bw = 0
\]
\begin{equation}
 \bJ  \bSigma+ \bSigma^T  \bJ^T + \frac{ \bD(\langle \bs \rangle ) }{\Omega}=0.
\label{variance2}
\end{equation}
By solving the above two equations one can estimate the mean values of the concentrations and their covariance matrix.

\section{Definition of stochastic focusing}
Stochastic focusing (SF) that we have discussed here has minor conceptual differences from the one described in Paulsson et al.~\cite{Paulsson2000}.
Consider the following reactions:
\begin{equation}
\xrightarrow{~~c_s~~} S \xrightarrow{\gamma_s s} \o,~~\xrightarrow{v(s) = 1/(K_M + s)}P \xrightarrow{\gamma_p p} \o,
\end{equation}
with $c_s$ a constant synthesis rate of $S$ and $v(s)$ a synthesis rate of $P$. The SF that we could discuss for the above system is independent of how fast the concentration $s$ fluctuates.
In \cite{Paulsson2000}, SF is however the only effect of rapid fluctuations in concentrations. This difference is due to the fact that the focus  in Paulsson et al.~\cite{Paulsson2000} was on what is the most probable state of $p$, and we focus on the mean value of $p$, more precisely $\langle v \rangle$ since the flux balance at the stationary state leads to $\langle v \rangle = \gamma_p \langle p \rangle$, resulting in $\langle v \rangle$ being proportional to $\langle p \rangle$.    To understand this difference, we need to understand the dynamics of $p$ that is correlated with the fluctuation in $s$.  When all reaction rates $c_s$, $\gamma_s s$, $v(s)$, and $\gamma_p p$ are in the same order of magnitude, $s$ and $p$ fluctuate on similar time scales.  If $s$ hits zero, the inhibition on $p$ is removed and thus the number of $p$ can rapidly increase to a very large number.  When $s$ increases to $1$, however,  the inhibition on $p$ appears and  the number of $p$ rapidly decreases.  Therefore, the time series profile of $p$ shows a flat lower bound at zero with many large sharp spikes.  This time series profile changes as the synthesis and degradation of $s$ become faster.  If the parameter values of $c_s$ and $\gamma_s$ increase such that  $s$ fluctuates much faster than $p$,  $p$ sees the averaged behavior of $s$ and is unlikely  to hit zero.  Thus, the strong/weak inhibition by $s$ is averaged out.   Due to this averaging, the sensitivity enhancement, i.e., SF, manifests itself.  This is why the SF was claimed to be the effect of a rapidly fluctuating $s$ \cite{Paulsson2000}.  However if one can observe the time series profile for a sufficiently long time to obtain good statistics of the spike heights,  the mean value of $p$ becomes actually independent from  how fast $c_s$ and $\gamma_s$ are, if the ratio $\gamma_s/c_s$ is presumed to be kept constant.  This means that the mean rates of $\gamma_p p$ and $v(s)$ also becomes time-scale independent. This is why the SF defined here becomes time-scale independent.



\begin{thebibliography}{50}%
\makeatletter
\providecommand \@ifxundefined [1]{%
 \@ifx{#1\undefined}
}%
\providecommand \@ifnum [1]{%
 \ifnum #1\expandafter \@firstoftwo
 \else \expandafter \@secondoftwo
 \fi
}%
\providecommand \@ifx [1]{%
 \ifx #1\expandafter \@firstoftwo
 \else \expandafter \@secondoftwo
 \fi
}%
\providecommand \natexlab [1]{#1}%
\providecommand \enquote  [1]{``#1''}%
\providecommand \bibnamefont  [1]{#1}%
\providecommand \bibfnamefont [1]{#1}%
\providecommand \citenamefont [1]{#1}%
\providecommand \href@noop [0]{\@secondoftwo}%
\providecommand \href [0]{\begingroup \@sanitize@url \@href}%
\providecommand \@href[1]{\@@startlink{#1}\@@href}%
\providecommand \@@href[1]{\endgroup#1\@@endlink}%
\providecommand \@sanitize@url [0]{\catcode `\\12\catcode `\$12\catcode
  `\&12\catcode `\#12\catcode `\^12\catcode `\_12\catcode `\%12\relax}%
\providecommand \@@startlink[1]{}%
\providecommand \@@endlink[0]{}%
\providecommand \url  [0]{\begingroup\@sanitize@url \@url }%
\providecommand \@url [1]{\endgroup\@href {#1}{\urlprefix }}%
\providecommand \urlprefix  [0]{URL }%
\providecommand \Eprint [0]{\href }%
\providecommand \doibase [0]{http://dx.doi.org/}%
\providecommand \selectlanguage [0]{\@gobble}%
\providecommand \bibinfo  [0]{\@secondoftwo}%
\providecommand \bibfield  [0]{\@secondoftwo}%
\providecommand \translation [1]{[#1]}%
\providecommand \BibitemOpen [0]{}%
\providecommand \bibitemStop [0]{}%
\providecommand \bibitemNoStop [0]{.\EOS\space}%
\providecommand \EOS [0]{\spacefactor3000\relax}%
\providecommand \BibitemShut  [1]{\csname bibitem#1\endcsname}%
\let\auto@bib@innerbib\@empty
\bibitem [{\citenamefont {Elowitz}\ \emph {et~al.}(2002)\citenamefont
  {Elowitz}, \citenamefont {Levine}, \citenamefont {Siggia},\ and\
  \citenamefont {Swain}}]{Elowitz2002}%
  \BibitemOpen
  \bibfield  {author} {\bibinfo {author} {\bibfnamefont {M.~B.}\ \bibnamefont
  {Elowitz}}, \bibinfo {author} {\bibfnamefont {A.~J.}\ \bibnamefont {Levine}},
  \bibinfo {author} {\bibfnamefont {E.~D.}\ \bibnamefont {Siggia}}, \ and\
  \bibinfo {author} {\bibfnamefont {P.~S.}\ \bibnamefont {Swain}},\ }\href
  {\doibase 10.1126/science.1070919} {\bibfield  {journal} {\bibinfo  {journal}
  {Science}\ }\textbf {\bibinfo {volume} {297}},\ \bibinfo {pages} {1183}
  (\bibinfo {year} {2002})}\BibitemShut {NoStop}%
\bibitem [{\citenamefont {Pedraza}\ and\ \citenamefont {van
  Oudenaarden}(2005)}]{Pedraza2005}%
  \BibitemOpen
  \bibfield  {author} {\bibinfo {author} {\bibfnamefont {J.~M.}\ \bibnamefont
  {Pedraza}}\ and\ \bibinfo {author} {\bibfnamefont {A.}~\bibnamefont {van
  Oudenaarden}},\ }\href {\doibase 10.1126/science.1109090} {\bibfield
  {journal} {\bibinfo  {journal} {Science}\ }\textbf {\bibinfo {volume}
  {307}},\ \bibinfo {pages} {1965} (\bibinfo {year} {2005})}\BibitemShut
  {NoStop}%
\bibitem [{\citenamefont {Xie}\ \emph {et~al.}(2008)\citenamefont {Xie},
  \citenamefont {Choi}, \citenamefont {Li}, \citenamefont {Lee},\ and\
  \citenamefont {Lia}}]{Xie2008a}%
  \BibitemOpen
  \bibfield  {author} {\bibinfo {author} {\bibfnamefont {X.~S.}\ \bibnamefont
  {Xie}}, \bibinfo {author} {\bibfnamefont {P.~J.}\ \bibnamefont {Choi}},
  \bibinfo {author} {\bibfnamefont {G.-W.}\ \bibnamefont {Li}}, \bibinfo
  {author} {\bibfnamefont {N.~K.}\ \bibnamefont {Lee}}, \ and\ \bibinfo
  {author} {\bibfnamefont {G.}~\bibnamefont {Lia}},\ }\href {\doibase
  10.1146/annurev.biophys.37.092607.174640} {\bibfield  {journal} {\bibinfo
  {journal} {Annu. Rev. Biophys.}\ }\textbf {\bibinfo {volume} {37}},\ \bibinfo
  {pages} {417} (\bibinfo {year} {2008})}\BibitemShut {NoStop}%
\bibitem [{\citenamefont {Yu}\ \emph {et~al.}(2006)\citenamefont {Yu},
  \citenamefont {Xiao}, \citenamefont {Ren}, \citenamefont {Lao},\ and\
  \citenamefont {Xie}}]{Yu2006}%
  \BibitemOpen
  \bibfield  {author} {\bibinfo {author} {\bibfnamefont {J.}~\bibnamefont
  {Yu}}, \bibinfo {author} {\bibfnamefont {J.}~\bibnamefont {Xiao}}, \bibinfo
  {author} {\bibfnamefont {X.}~\bibnamefont {Ren}}, \bibinfo {author}
  {\bibfnamefont {K.}~\bibnamefont {Lao}}, \ and\ \bibinfo {author}
  {\bibfnamefont {X.~S.}\ \bibnamefont {Xie}},\ }\href {\doibase
  10.1126/science.1119623} {\bibfield  {journal} {\bibinfo  {journal}
  {Science}\ }\textbf {\bibinfo {volume} {311}},\ \bibinfo {pages} {1600}
  (\bibinfo {year} {2006})}\BibitemShut {NoStop}%
\bibitem [{\citenamefont {Oehler}\ \emph {et~al.}(1994)\citenamefont {Oehler},
  \citenamefont {Amouyal}, \citenamefont {Kolkhof}, \citenamefont {von
  Wilcken-Bergmann},\ and\ \citenamefont {M{\"u}ller-Hill}}]{Oehler1994}%
  \BibitemOpen
  \bibfield  {author} {\bibinfo {author} {\bibfnamefont {S.}~\bibnamefont
  {Oehler}}, \bibinfo {author} {\bibfnamefont {M.}~\bibnamefont {Amouyal}},
  \bibinfo {author} {\bibfnamefont {P.}~\bibnamefont {Kolkhof}}, \bibinfo
  {author} {\bibfnamefont {B.}~\bibnamefont {von Wilcken-Bergmann}}, \ and\
  \bibinfo {author} {\bibfnamefont {B.}~\bibnamefont {M{\"u}ller-Hill}},\
  }\href@noop {} {\bibfield  {journal} {\bibinfo  {journal} {EMBO J.}\ }\textbf
  {\bibinfo {volume} {13}},\ \bibinfo {pages} {3348} (\bibinfo {year}
  {1994})}\BibitemShut {NoStop}%
\bibitem [{\citenamefont {Choi}\ \emph {et~al.}(2008)\citenamefont {Choi},
  \citenamefont {Cai}, \citenamefont {Frieda},\ and\ \citenamefont
  {Xie}}]{Choi2008}%
  \BibitemOpen
  \bibfield  {author} {\bibinfo {author} {\bibfnamefont {P.~J.}\ \bibnamefont
  {Choi}}, \bibinfo {author} {\bibfnamefont {L.}~\bibnamefont {Cai}}, \bibinfo
  {author} {\bibfnamefont {K.}~\bibnamefont {Frieda}}, \ and\ \bibinfo {author}
  {\bibfnamefont {X.~S.}\ \bibnamefont {Xie}},\ }\href
  {http://wuos.org/content/322/5900/442.short} {\bibfield  {journal} {\bibinfo
  {journal} {Science}\ }\textbf {\bibinfo {volume} {322}},\ \bibinfo {pages}
  {442} (\bibinfo {year} {2008})}\BibitemShut {NoStop}%
\bibitem [{\citenamefont {Qian}(2012)}]{Qian2012}%
  \BibitemOpen
  \bibfield  {author} {\bibinfo {author} {\bibfnamefont {H.}\ \bibnamefont
  {Qian}},\ }\href@noop {} {\bibfield  {journal} {\bibinfo  {journal}
  {Annu. Rev. Biophys.}\ }\textbf {\bibinfo {volume} {41}},\ \bibinfo {pages} {179}
  (\bibinfo {year} {2012})}\BibitemShut {NoStop}%
\bibitem [{\citenamefont {Gillespie}(1992)}]{Gillespie1992}%
  \BibitemOpen
  \bibfield  {author} {\bibinfo {author} {\bibfnamefont {D.~T.}\ \bibnamefont
  {Gillespie}},\ }\href@noop {} {\bibfield  {journal} {\bibinfo  {journal}
  {Physica A}\ }\textbf {\bibinfo {volume} {188}},\ \bibinfo {pages} {404}
  (\bibinfo {year} {1992})}\BibitemShut {NoStop}%
\bibitem [{\citenamefont {McQuarrie}(1967)}]{Mcquarrie1967}%
  \BibitemOpen
  \bibfield  {author} {\bibinfo {author} {\bibfnamefont {D.~A.}\ \bibnamefont
  {McQuarrie}},\ }\href@noop {} {\bibfield  {journal} {\bibinfo  {journal} {J.
  Appl. Probab.}\ }\textbf {\bibinfo {volume} {4}},\ \bibinfo {pages} {413}
  (\bibinfo {year} {1967})}\BibitemShut {NoStop}%
\bibitem [{\citenamefont {Liang}\ and\ \citenamefont {Qian}(2010)}]{Liang2010}%
  \BibitemOpen
  \bibfield  {author} {\bibinfo {author} {\bibfnamefont {J.}~\bibnamefont
  {Liang}}\ and\ \bibinfo {author} {\bibfnamefont {H.}~\bibnamefont {Qian}},\
  }\href {\doibase 10.1007/s11390-010-9312-6} {\bibfield  {journal} {\bibinfo
  {journal} {J. Comput. Sci. Technol.}\ }\textbf {\bibinfo
  {volume} {25}},\ \bibinfo {pages} {154} (\bibinfo {year} {2010})}\BibitemShut
  {NoStop}%
\bibitem [{\citenamefont {Gillespie}(1977)}]{Gillespie1977}%
  \BibitemOpen
  \bibfield  {author} {\bibinfo {author} {\bibfnamefont {D.~T.}\ \bibnamefont
  {Gillespie}},\ }\href@noop {} {\bibfield  {journal} {\bibinfo  {journal} {J.
  Phys. Chem.}\ }\textbf {\bibinfo {volume} {81}},\ \bibinfo {pages} {2340}
  (\bibinfo {year} {1977})}\BibitemShut {NoStop}%
\bibitem [{\citenamefont {{Van Kampen}}(2001)}]{Kampen2001}%
  \BibitemOpen
  \bibfield  {author} {\bibinfo {author} {\bibfnamefont {N.~G.}\ \bibnamefont
  {{Van Kampen}}},\ }\href@noop {} {\emph {\bibinfo {title} {{Stochastic
  Processes in Physics and Chemistry.}}}},\ \bibinfo {edition} {3rd}\ ed.\
  (\bibinfo  {publisher} {North Holland},\ \bibinfo {year} {2001})\BibitemShut
  {NoStop}%
\bibitem [{\citenamefont {G\'{o}mez-Uribe}\ and\ \citenamefont
  {Verghese}(2007)}]{Gomez2007}%
  \BibitemOpen
  \bibfield  {author} {\bibinfo {author} {\bibfnamefont {C.~A.}\ \bibnamefont
  {G\'{o}mez-Uribe}}\ and\ \bibinfo {author} {\bibfnamefont {G.~C.}\
  \bibnamefont {Verghese}},\ }\href {\doibase 10.1063/1.2408422} {\bibfield
  {journal} {\bibinfo  {journal} {J. Chem. Phys.}\ }\textbf {\bibinfo {volume}
  {126}},\ \bibinfo {pages} {24109} (\bibinfo {year} {2007})}\BibitemShut
  {NoStop}%
\bibitem [{\citenamefont {Scott}, \citenamefont {Hwa},\ and\ \citenamefont
  {Ingalls}(2007)}]{Scott2007}%
  \BibitemOpen
  \bibfield  {author} {\bibinfo {author} {\bibfnamefont {M.}~\bibnamefont
  {Scott}}, \bibinfo {author} {\bibfnamefont {T.}~\bibnamefont {Hwa}}, \ and\
  \bibinfo {author} {\bibfnamefont {B.}~\bibnamefont {Ingalls}},\ }\href
  {\doibase 10.1073/pnas.0610468104} {\bibfield  {journal} {\bibinfo  {journal}
  {Proc. Natl. Acad. Sci. U.S.A.}\ }\textbf {\bibinfo {volume} {104}},\
  \bibinfo {pages} {7402} (\bibinfo {year} {2007})}\BibitemShut {NoStop}%
\bibitem [{\citenamefont {Munsky}\ and\ \citenamefont
  {Khammash}(2006)}]{Munsky2006}%
  \BibitemOpen
  \bibfield  {author} {\bibinfo {author} {\bibfnamefont {B.}~\bibnamefont
  {Munsky}}\ and\ \bibinfo {author} {\bibfnamefont {M.}~\bibnamefont
  {Khammash}},\ }\href {\doibase 10.1063/1.2145882} {\bibfield  {journal}
  {\bibinfo  {journal} {J. Chem. Phys.}\ }\textbf {\bibinfo {volume} {124}},\
  \bibinfo {pages} {44104} (\bibinfo {year} {2006})}\BibitemShut {NoStop}%
\bibitem [{\citenamefont {Kim}\ and\ \citenamefont {Price}(2010)}]{KimPJ2010}%
  \BibitemOpen
  \bibfield  {author} {\bibinfo {author} {\bibfnamefont {P.-J.}\ \bibnamefont
  {Kim}}\ and\ \bibinfo {author} {\bibfnamefont {N.~D.}\ \bibnamefont
  {Price}},\ }\href {\doibase 10.1103/PhysRevLett.104.148103} {\bibfield
  {journal} {\bibinfo  {journal} {Phys. Rev. Lett.}\ }\textbf {\bibinfo
  {volume} {104}},\ \bibinfo {pages} {9} (\bibinfo {year} {2010})}\BibitemShut
  {NoStop}%
\bibitem [{\citenamefont {Fell}(1992)}]{Fell1992}%
  \BibitemOpen
  \bibfield  {author} {\bibinfo {author} {\bibfnamefont {D.~A.}\ \bibnamefont
  {Fell}},\ }\href@noop {} {\bibfield  {journal} {\bibinfo  {journal} {Biochem.
  J.}\ }\textbf {\bibinfo {volume} {286}},\ \bibinfo {pages} {313} (\bibinfo
  {year} {1992})}\BibitemShut {NoStop}%
\bibitem [{\citenamefont {Kacser}\ and\ \citenamefont
  {Burns}(1995)}]{Kacser1995}%
  \BibitemOpen
  \bibfield  {author} {\bibinfo {author} {\bibfnamefont {H.}~\bibnamefont
  {Kacser}}\ and\ \bibinfo {author} {\bibfnamefont {J.~A.}\ \bibnamefont
  {Burns}},\ }\href@noop {} {\bibfield  {journal} {\bibinfo  {journal}
  {Biochem. Soc. Trans.}\ }\textbf {\bibinfo {volume} {23}},\ \bibinfo {pages}
  {341} (\bibinfo {year} {1995})}\BibitemShut {NoStop}%
\bibitem [{\citenamefont {Fell}(1996)}]{Fell1996}%
  \BibitemOpen
  \bibfield  {author} {\bibinfo {author} {\bibfnamefont {D.~A.}\ \bibnamefont
  {Fell}},\ }\href@noop {} {\emph {\bibinfo {title} {{Understanding the Control
  of Metabolism.}}}}\ (\bibinfo  {publisher} {London, Portland Press},\
  \bibinfo {year} {1996})\BibitemShut {NoStop}%
\bibitem [{\citenamefont {Paulsson}(2004)}]{Paulsson2004}%
  \BibitemOpen
  \bibfield  {author} {\bibinfo {author} {\bibfnamefont {J.}~\bibnamefont
  {Paulsson}},\ }\href {\doibase 10.1038/nature02257} {\bibfield  {journal}
  {\bibinfo  {journal} {Nature}\ }\textbf {\bibinfo {volume} {427}},\ \bibinfo
  {pages} {415} (\bibinfo {year} {2004})}\BibitemShut {NoStop}%
\bibitem [{\citenamefont {Sauro}(2011)}]{Sauro2011}%
  \BibitemOpen
  \bibfield  {author} {\bibinfo {author} {\bibfnamefont {H.~M.}\ \bibnamefont
  {Sauro}},\ }\href@noop {} {\emph {\bibinfo {title} {{Enzyme Kinetics for
  Systems Biology}}}},\ \bibinfo {edition} {2nd}\ ed.\ (\bibinfo  {publisher}
  {Ambrosius Publishing and Future Skill Software},\ \bibinfo {year}
  {2012})\BibitemShut {NoStop}%
\bibitem [{\citenamefont {Hooshangi}, \citenamefont {Thiberge},\ and\
  \citenamefont {Weiss}(2005)}]{Hooshangi2005}%
  \BibitemOpen
  \bibfield  {author} {\bibinfo {author} {\bibfnamefont {S.}~\bibnamefont
  {Hooshangi}}, \bibinfo {author} {\bibfnamefont {S.}~\bibnamefont {Thiberge}},
  \ and\ \bibinfo {author} {\bibfnamefont {R.}~\bibnamefont {Weiss}},\ }\href
  {\doibase 10.1073/pnas.0408507102} {\bibfield  {journal} {\bibinfo  {journal}
  {Proc. Natl. Acad. Sci. U.S.A.}\ }\textbf {\bibinfo {volume} {102}},\
  \bibinfo {pages} {3581} (\bibinfo {year} {2005})}\BibitemShut {NoStop}%
\bibitem [{\citenamefont {T{\u{a}}nase-Nicola}, \citenamefont {Warren},\ and\
  \citenamefont {ten Wolde}(2006)}]{Tanase2006}%
  \BibitemOpen
  \bibfield  {author} {\bibinfo {author} {\bibfnamefont {S.}~\bibnamefont
  {T{\u{a}}nase-Nicola}}, \bibinfo {author} {\bibfnamefont {P.~B.}\
  \bibnamefont {Warren}}, \ and\ \bibinfo {author} {\bibfnamefont {P.~R.}\
  \bibnamefont {ten Wolde}},\ }\href@noop {} {\bibfield  {journal} {\bibinfo
  {journal} {Phys. Rev. Lett.}\ }\textbf {\bibinfo {volume} {97}},\ \bibinfo
  {pages} {68102} (\bibinfo {year} {2006})}\BibitemShut {NoStop}%
\bibitem [{\citenamefont {Bruggeman}, \citenamefont {Bl{\"u}thgen},\ and\
  \citenamefont {Westerhoff}(2009)}]{Bruggeman2009}%
  \BibitemOpen
  \bibfield  {author} {\bibinfo {author} {\bibfnamefont {F.}~\bibnamefont
  {Bruggeman}}, \bibinfo {author} {\bibfnamefont {N.}~\bibnamefont
  {Bl{\"u}thgen}}, \ and\ \bibinfo {author} {\bibfnamefont {H.}~\bibnamefont
  {Westerhoff}},\ }\href {\doibase 10.1371/Citation} {\bibfield  {journal}
  {\bibinfo  {journal} {PLoS Comput. Biol.}\ }\textbf {\bibinfo {volume} {5}},\
  \bibinfo {pages} {e1000506} (\bibinfo {year} {2009})}\BibitemShut {NoStop}%
\bibitem [{\citenamefont {Paulsson}, \citenamefont {Berg},\ and\ \citenamefont
  {Ehrenberg}(2000)}]{Paulsson2000}%
  \BibitemOpen
  \bibfield  {author} {\bibinfo {author} {\bibfnamefont {J.}~\bibnamefont
  {Paulsson}}, \bibinfo {author} {\bibfnamefont {O.~G.}\ \bibnamefont {Berg}},
  \ and\ \bibinfo {author} {\bibfnamefont {M.}~\bibnamefont {Ehrenberg}},\
  }\href {\doibase 10.1073/pnas.110057697} {\bibfield  {journal} {\bibinfo
  {journal} {Proc. Natl. Acad. Sci. U. S. A.}\ }\textbf {\bibinfo {volume}
  {97}},\ \bibinfo {pages} {7148} (\bibinfo {year} {2000})}\BibitemShut
  {NoStop}%
\bibitem [{\citenamefont {Qian}(2011)}]{Qian2011}%
  \BibitemOpen
  \bibfield  {author} {\bibinfo {author} {\bibfnamefont {H.}~\bibnamefont
  {Qian}},\ }\href {\doibase 10.1088/0951-7715/24/6/R01} {\bibfield  {journal}
  {\bibinfo  {journal} {Nonlinearity}\ }\textbf {\bibinfo {volume} {24}},\
  \bibinfo {pages} {R19} (\bibinfo {year} {2011})}\BibitemShut {NoStop}%
\bibitem [{\citenamefont {Ochab-Marcinek}\ and\ \citenamefont
  {Tabaka}(2010)}]{Ochab2010}%
  \BibitemOpen
  \bibfield  {author} {\bibinfo {author} {\bibfnamefont {A.}~\bibnamefont
  {Ochab-Marcinek}}\ and\ \bibinfo {author} {\bibfnamefont {M.}~\bibnamefont
  {Tabaka}},\ }\href {\doibase
  10.1073/pnas.1008965107/-/DCSupplemental.www.pnas.org/cgi/doi/10.1073/pnas.1008965107}
  {\bibfield  {journal} {\bibinfo  {journal} {Proc. Natl. Acad. Sci. U. S. A.}\
  }\textbf {\bibinfo {volume} {107}},\ \bibinfo {pages} {22096} (\bibinfo
  {year} {2010})}\BibitemShut {NoStop}%
\bibitem [{\citenamefont {Taniguchi}\ \emph {et~al.}(2010)\citenamefont
  {Taniguchi}, \citenamefont {Choi}, \citenamefont {Li}, \citenamefont {Chen},
  \citenamefont {Babu}, \citenamefont {Hearn}, \citenamefont {Emili},\ and\
  \citenamefont {Xie}}]{Taniguchi2010}%
  \BibitemOpen
  \bibfield  {author} {\bibinfo {author} {\bibfnamefont {Y.}~\bibnamefont
  {Taniguchi}}, \bibinfo {author} {\bibfnamefont {P.~J.}\ \bibnamefont {Choi}},
  \bibinfo {author} {\bibfnamefont {G.-W.}\ \bibnamefont {Li}}, \bibinfo
  {author} {\bibfnamefont {H.}~\bibnamefont {Chen}}, \bibinfo {author}
  {\bibfnamefont {M.}~\bibnamefont {Babu}}, \bibinfo {author} {\bibfnamefont
  {J.}~\bibnamefont {Hearn}}, \bibinfo {author} {\bibfnamefont
  {A.}~\bibnamefont {Emili}}, \ and\ \bibinfo {author} {\bibfnamefont {X.~S.}\
  \bibnamefont {Xie}},\ }\href {\doibase 10.1126/science.1188308} {\bibfield
  {journal} {\bibinfo  {journal} {Science}\ }\textbf {\bibinfo {volume}
  {329}},\ \bibinfo {pages} {533} (\bibinfo {year} {2010})}\BibitemShut
  {NoStop}%
\bibitem [{\citenamefont {Kim}\ and\ \citenamefont {Sauro}(2012)}]{Kim2012}%
  \BibitemOpen
  \bibfield  {author} {\bibinfo {author} {\bibfnamefont {K.~H.}\ \bibnamefont
  {Kim}}\ and\ \bibinfo {author} {\bibfnamefont {H.~M.}\ \bibnamefont
  {Sauro}},\ }\href {\doibase 10.1371/journal.pcbi.1002344} {\bibfield
  {journal} {\bibinfo  {journal} {PLoS Comput. Biol.}\ }\textbf
  {\bibinfo {volume} {8}},\ \bibinfo {pages} {e1002344} (\bibinfo {year}
  {2012})}\BibitemShut {NoStop}%
\bibitem [{\citenamefont {Entus}, \citenamefont {Aufderheide},\ and\
  \citenamefont {Sauro}(2007)}]{Entus2007}%
  \BibitemOpen
  \bibfield  {author} {\bibinfo {author} {\bibfnamefont {R.}~\bibnamefont
  {Entus}}, \bibinfo {author} {\bibfnamefont {B.}~\bibnamefont {Aufderheide}},
  \ and\ \bibinfo {author} {\bibfnamefont {H.~M.}\ \bibnamefont {Sauro}},\
  }\href@noop {} {\bibfield  {journal} {\bibinfo  {journal} {Syst. Synth.
  Biol.}\ }\textbf {\bibinfo {volume} {1}},\ \bibinfo {pages} {119} (\bibinfo
  {year} {2007})}\BibitemShut {NoStop}%
\bibitem [{\citenamefont {Kaplan}\ \emph {et~al.}(2008)\citenamefont {Kaplan},
  \citenamefont {Bren}, \citenamefont {Dekel},\ and\ \citenamefont
  {Alon}}]{Kaplan2008}%
  \BibitemOpen
  \bibfield  {author} {\bibinfo {author} {\bibfnamefont {S.}~\bibnamefont
  {Kaplan}}, \bibinfo {author} {\bibfnamefont {A.}~\bibnamefont {Bren}},
  \bibinfo {author} {\bibfnamefont {E.}~\bibnamefont {Dekel}}, \ and\ \bibinfo
  {author} {\bibfnamefont {U.}~\bibnamefont {Alon}},\ }\href {\doibase
  10.1038/msb.2008.43} {\bibfield  {journal} {\bibinfo  {journal} {Mol. Syst.
  Biol.}\ }\textbf {\bibinfo {volume} {4}},\ \bibinfo {pages} {203} (\bibinfo
  {year} {2008})}\BibitemShut {NoStop}%
\bibitem [{\citenamefont {Ozbudak}\ \emph {et~al.}(2004)\citenamefont
  {Ozbudak}, \citenamefont {Thattai}, \citenamefont {Lim}, \citenamefont
  {Shraiman},\ and\ \citenamefont {{Van Oudenaarden}}}]{Ozbudak2004}%
  \BibitemOpen
  \bibfield  {author} {\bibinfo {author} {\bibfnamefont {E.~M.}\ \bibnamefont
  {Ozbudak}}, \bibinfo {author} {\bibfnamefont {M.}~\bibnamefont {Thattai}},
  \bibinfo {author} {\bibfnamefont {H.~N.}\ \bibnamefont {Lim}}, \bibinfo
  {author} {\bibfnamefont {B.~I.}\ \bibnamefont {Shraiman}}, \ and\ \bibinfo
  {author} {\bibfnamefont {A.}~\bibnamefont {{Van Oudenaarden}}},\ }\href
  {\doibase 10.1038/nature02298} {\bibfield  {journal} {\bibinfo  {journal}
  {Nature}\ }\textbf {\bibinfo {volume} {427}},\ \bibinfo {pages} {737}
  (\bibinfo {year} {2004})}\BibitemShut {NoStop}%
\bibitem [{\citenamefont {Brandman}\ \emph {et~al.}(2005)\citenamefont
  {Brandman}, \citenamefont {Ferrell}, \citenamefont {Li},\ and\ \citenamefont
  {Meyer}}]{Brandman2005}%
  \BibitemOpen
  \bibfield  {author} {\bibinfo {author} {\bibfnamefont {O.}~\bibnamefont
  {Brandman}}, \bibinfo {author} {\bibfnamefont {J.~E.}\ \bibnamefont
  {Ferrell,~Jr.}}, \bibinfo {author} {\bibfnamefont {R.}~\bibnamefont {Li}}, \ and\
  \bibinfo {author} {\bibfnamefont {T.}~\bibnamefont {Meyer}},\ }\href
  {\doibase 10.1126/science.1113834} {\bibfield  {journal} {\bibinfo  {journal}
  {Science}\ }\textbf {\bibinfo {volume} {310}},\ \bibinfo
  {pages} {496} (\bibinfo {year} {2005})}\BibitemShut {NoStop}%
\bibitem [{\citenamefont {Sauro}(2009)}]{Sauro2009}%
  \BibitemOpen
  \bibfield  {author} {\bibinfo {author} {\bibfnamefont {H.~M.}\ \bibnamefont
  {Sauro}},\ }in\ \href@noop {} {\emph {\bibinfo {booktitle} {Computational
  Systems Biology}}}\ (\bibinfo  {publisher} {Humana Press},\ \bibinfo
  {address} {New York},\ \bibinfo {year} {2009})\ Chap.~\bibinfo {chapter}
  {13}, pp.\ \bibinfo {pages} {269--309}\BibitemShut {NoStop}%
\bibitem [{\citenamefont {Yu}\ \emph {et~al.}(2008)\citenamefont {Yu},
  \citenamefont {Pesce}, \citenamefont {Colman-Lerner}, \citenamefont {Lok},
  \citenamefont {Pincus}, \citenamefont {Serra}, \citenamefont {Holl},
  \citenamefont {Benjamin}, \citenamefont {Gordon},\ and\ \citenamefont
  {Brent}}]{Yu2008}%
  \BibitemOpen
  \bibfield  {author} {\bibinfo {author} {\bibfnamefont {R.~C.}\ \bibnamefont
  {Yu}}, \bibinfo {author} {\bibfnamefont {C.~G.}\ \bibnamefont {Pesce}},
  \bibinfo {author} {\bibfnamefont {A.}~\bibnamefont {Colman-Lerner}}, \bibinfo
  {author} {\bibfnamefont {L.}~\bibnamefont {Lok}}, \bibinfo {author}
  {\bibfnamefont {D.}~\bibnamefont {Pincus}}, \bibinfo {author} {\bibfnamefont
  {E.}~\bibnamefont {Serra}}, \bibinfo {author} {\bibfnamefont
  {M.}~\bibnamefont {Holl}}, \bibinfo {author} {\bibfnamefont {K.}~\bibnamefont
  {Benjamin}}, \bibinfo {author} {\bibfnamefont {A.}~\bibnamefont {Gordon}}, \
  and\ \bibinfo {author} {\bibfnamefont {R.}~\bibnamefont {Brent}},\ }\href
  {\doibase 10.1038/nature07513} {\bibfield  {journal} {\bibinfo  {journal}
  {Nature}\ }\textbf {\bibinfo {volume} {456}},\ \bibinfo {pages} {755}
  (\bibinfo {year} {2008})}\BibitemShut {NoStop}%
\bibitem [{\citenamefont {Sturm}\ \emph {et~al.}(2010)\citenamefont {Sturm},
  \citenamefont {Orton}, \citenamefont {Grindlay}, \citenamefont {Birtwistle},
  \citenamefont {Vyshemirsky}, \citenamefont {Gilbert}, \citenamefont {Calder},
  \citenamefont {Pitt}, \citenamefont {Kholodenko},\ and\ \citenamefont
  {Kolch}}]{Sturm2010}%
  \BibitemOpen
  \bibfield  {author} {\bibinfo {author} {\bibfnamefont {O.~E.}\ \bibnamefont
  {Sturm}}, \bibinfo {author} {\bibfnamefont {R.}~\bibnamefont {Orton}},
  \bibinfo {author} {\bibfnamefont {J.}~\bibnamefont {Grindlay}}, \bibinfo
  {author} {\bibfnamefont {M.}~\bibnamefont {Birtwistle}}, \bibinfo {author}
  {\bibfnamefont {V.}~\bibnamefont {Vyshemirsky}}, \bibinfo {author}
  {\bibfnamefont {D.}~\bibnamefont {Gilbert}}, \bibinfo {author} {\bibfnamefont
  {M.}~\bibnamefont {Calder}}, \bibinfo {author} {\bibfnamefont
  {A.}~\bibnamefont {Pitt}}, \bibinfo {author} {\bibfnamefont {B.}~\bibnamefont
  {Kholodenko}}, \ and\ \bibinfo {author} {\bibfnamefont {W.}~\bibnamefont
  {Kolch}},\ }\href {\doibase 10.1126/scisignal.2001212} {\bibfield  {journal}
  {\bibinfo  {journal} {Sci. Signal.}\ }\textbf {\bibinfo {volume} {3}},\
  \bibinfo {pages} {ra90} (\bibinfo {year} {2010})}\BibitemShut {NoStop}%
\bibitem [{\citenamefont {Sauro}\ and\ \citenamefont
  {Ingalls}(2007)}]{Sauro2007}%
  \BibitemOpen
  \bibfield  {author} {\bibinfo {author} {\bibfnamefont {H.~M.}\ \bibnamefont
  {Sauro}}\ and\ \bibinfo {author} {\bibfnamefont {B.}~\bibnamefont
  {Ingalls}},\ }\href@noop {} {\bibfield  {journal} {\bibinfo  {journal}
  {arXiv:0710.5195[q-bio.MN]}\ } (\bibinfo {year} {2007})}
  \BibitemShut {NoStop}%
\bibitem [{\citenamefont {Thattai}\ and\ \citenamefont {van
  Oudenaarden}(2002)}]{Thattai2002}%
  \BibitemOpen
  \bibfield  {author} {\bibinfo {author} {\bibfnamefont {M.}~\bibnamefont
  {Thattai}}\ and\ \bibinfo {author} {\bibfnamefont {A.}~\bibnamefont {van
  Oudenaarden}},\ }\href {\doibase 10.1016/S0006-3495(02)75635-X} {\bibfield
  {journal} {\bibinfo  {journal} {Biophys. J.}\ }\textbf {\bibinfo
  {volume} {82}},\ \bibinfo {pages} {2943} (\bibinfo {year}
  {2002})}\BibitemShut {NoStop}%
\bibitem [{\citenamefont {Birtwistle}\ \emph
  {et~al.}(2012{\natexlab{a}})\citenamefont {Birtwistle}, \citenamefont
  {Rauch}, \citenamefont {Kiyatkin}, \citenamefont {Aksamitiene}, \citenamefont
  {{Dobrzynski}}, \citenamefont {Hoek}, \citenamefont {Kolch},
  \citenamefont {Ogunnaike},\ and\ \citenamefont
  {Kholodenko}}]{Birtwistle2012}%
  \BibitemOpen
  \bibfield  {author} {\bibinfo {author} {\bibfnamefont {M.~R.}\ \bibnamefont
  {Birtwistle}}, \bibinfo {author} {\bibfnamefont {J.}~\bibnamefont {Rauch}},
  \bibinfo {author} {\bibfnamefont {A.}~\bibnamefont {Kiyatkin}}, \bibinfo
  {author} {\bibfnamefont {E.}~\bibnamefont {Aksamitiene}}, \bibinfo {author}
  {\bibfnamefont {M.}~\bibnamefont {{Dobrzy\'{n}ski}}}, \bibinfo {author}
  {\bibfnamefont {J.~B.}\ \bibnamefont {Hoek}}, \bibinfo {author}
  {\bibfnamefont {W.}~\bibnamefont {Kolch}}, \bibinfo {author} {\bibfnamefont
  {B.~A.}\ \bibnamefont {Ogunnaike}}, \ and\ \bibinfo {author} {\bibfnamefont
  {B.~N.}\ \bibnamefont {Kholodenko}},\ }\href {\doibase
  10.1186/1752-0509-6-109} {\bibfield  {journal} {\bibinfo  {journal} {BMC
  Syst. Biol.}\ }\textbf {\bibinfo {volume} {6}},\ \bibinfo {pages} {109}
  (\bibinfo {year} {2012}{\natexlab{a}})}\BibitemShut {NoStop}%
\bibitem [{\citenamefont {Gammaitoni}\ \emph {et~al.}(1998)\citenamefont
  {Gammaitoni}, \citenamefont {H{\"{a}}nggi}, \citenamefont {Jung},\ and\
  \citenamefont {Marchesoni}}]{Gammaitoni1998}%
  \BibitemOpen
  \bibfield  {author} {\bibinfo {author} {\bibfnamefont {L.}~\bibnamefont
  {Gammaitoni}}, \bibinfo {author} {\bibfnamefont {P.}~\bibnamefont
  {H{\"{a}}nggi}}, \bibinfo {author} {\bibfnamefont {P.}~\bibnamefont {Jung}},
  \ and\ \bibinfo {author} {\bibfnamefont {F.}~\bibnamefont {Marchesoni}},\
  }\href {http://eprints.lancs.ac.uk/32122/} {\bibfield  {journal} {\bibinfo
  {journal} {Rev. Mod. Phys.}\ }\textbf {\bibinfo {volume} {70}},\ \bibinfo
  {pages} {223} (\bibinfo {year} {1998})}\BibitemShut {NoStop}%
\bibitem [{\citenamefont {Paulsson}\ and\ \citenamefont
  {Ehrenberg}(2000)}]{Paulsson2000a}%
  \BibitemOpen
  \bibfield  {author} {\bibinfo {author} {\bibfnamefont {J.}~\bibnamefont
  {Paulsson}}\ and\ \bibinfo {author} {\bibfnamefont {M.}~\bibnamefont
  {Ehrenberg}},\ }\href {http://www.ncbi.nlm.nih.gov/pubmed/10990965}
  {\bibfield  {journal} {\bibinfo  {journal} {Phys. Rev. Lett.}\
  }\textbf {\bibinfo {volume} {84}},\ \bibinfo {pages} {5447} (\bibinfo {year}
  {2000})}\BibitemShut {NoStop}%
\bibitem [{\citenamefont {Friedman}, \citenamefont {Cai},\ and\ \citenamefont
  {Xie}(2006)}]{Friedman2006}%
  \BibitemOpen
  \bibfield  {author} {\bibinfo {author} {\bibfnamefont {N.}~\bibnamefont
  {Friedman}}, \bibinfo {author} {\bibfnamefont {L.}~\bibnamefont {Cai}}, \
  and\ \bibinfo {author} {\bibfnamefont {X.}~\bibnamefont {Xie}},\ }\href
  {\doibase 10.1103/PhysRevLett.97.168302} {\bibfield  {journal} {\bibinfo
  {journal} {Phys. Rev. Lett.}\ }\textbf {\bibinfo {volume} {97}},\
  \bibinfo {pages} {1} (\bibinfo {year} {2006})}\BibitemShut {NoStop}%
\bibitem [{\citenamefont {Shahrezaei}\ and\ \citenamefont
  {Swain}(2008)}]{Shahrezaei2008}%
  \BibitemOpen
  \bibfield  {author} {\bibinfo {author} {\bibfnamefont {V.}~\bibnamefont
  {Shahrezaei}}\ and\ \bibinfo {author} {\bibfnamefont {P.~S.}\ \bibnamefont
  {Swain}},\ }\href {\doibase 10.1073/pnas.0803850105} {\bibfield  {journal}
  {\bibinfo  {journal} {Proc. Natl. Acad. Sci. U. S. A.}\ }\textbf {\bibinfo {volume}
  {105}},\ \bibinfo {pages} {17256} (\bibinfo {year} {2008})}\BibitemShut
  {NoStop}%
\bibitem [{\citenamefont {Birtwistle}\ \emph
  {et~al.}(2012{\natexlab{b}})\citenamefont {Birtwistle}, \citenamefont {von
  Kriegsheim}, \citenamefont {Dobrzyński}, \citenamefont {Kholodenko},\ and\
  \citenamefont {Kolch}}]{Birtwistle2012a}%
  \BibitemOpen
  \bibfield  {author} {\bibinfo {author} {\bibfnamefont {M.~R.}\ \bibnamefont
  {Birtwistle}}, \bibinfo {author} {\bibfnamefont {A.}~\bibnamefont {von
  Kriegsheim}}, \bibinfo {author} {\bibfnamefont {M.}~\bibnamefont
  {Dobrzy{\'n}ski}}, \bibinfo {author} {\bibfnamefont {B.~N.}\ \bibnamefont
  {Kholodenko}}, \ and\ \bibinfo {author} {\bibfnamefont {W.}~\bibnamefont
  {Kolch}},\ }\href {\doibase 10.1039/c2mb25168j} {\bibfield  {journal}
  {\bibinfo  {journal} {Mol. Biosyst.}\ }\textbf {\bibinfo {volume}
  {8}},\ \bibinfo {pages} {3068} (\bibinfo {year}
  {2012}{\natexlab{b}})}\BibitemShut {NoStop}%
\bibitem [{\citenamefont {Arkin}, \citenamefont {Ross},\ and\ \citenamefont
  {McAdams}(1998)}]{Arkin1998}%
  \BibitemOpen
  \bibfield  {author} {\bibinfo {author} {\bibfnamefont {A.~P.}\ \bibnamefont
  {Arkin}}, \bibinfo {author} {\bibfnamefont {J.}~\bibnamefont {Ross}}, \ and\
  \bibinfo {author} {\bibfnamefont {H.~H.}\ \bibnamefont {McAdams}},\ }\href
  {http://www.pubmedcentral.nih.gov/articlerender.fcgi?artid=1460268&tool=pmcentrez&rendertype=abstract}
  {\bibfield  {journal} {\bibinfo  {journal} {Genetics}\ }\textbf {\bibinfo
  {volume} {149}},\ \bibinfo {pages} {1633} (\bibinfo {year}
  {1998})}\BibitemShut {NoStop}%
\bibitem [{\citenamefont {Maamar}, \citenamefont {Raj},\ and\ \citenamefont
  {Dubnau}(2007)}]{Maamar2007}%
  \BibitemOpen
  \bibfield  {author} {\bibinfo {author} {\bibfnamefont {H.}~\bibnamefont
  {Maamar}}, \bibinfo {author} {\bibfnamefont {A.}~\bibnamefont {Raj}}, \ and\
  \bibinfo {author} {\bibfnamefont {D.}~\bibnamefont {Dubnau}},\ }\href
  {\doibase 10.1126/science.1140818} {\bibfield  {journal} {\bibinfo  {journal}
  {Science}\ }\textbf {\bibinfo {volume} {317}},\ \bibinfo {pages} {526}
  (\bibinfo {year} {2007})}\BibitemShut {NoStop}%
\bibitem [{\citenamefont {Wang}\ and\ \citenamefont {Zhang}(2011)}]{Wang2011}%
  \BibitemOpen
  \bibfield  {author} {\bibinfo {author} {\bibfnamefont {Z.}~\bibnamefont
  {Wang}}\ and\ \bibinfo {author} {\bibfnamefont {J.}~\bibnamefont {Zhang}},\
  }\href {\doibase 10.1073/pnas.1100059108} {\bibfield  {journal} {\bibinfo
  {journal} {Proc. Natl. Acad. Sci. U. S. A.}\ }\textbf {\bibinfo {volume}
  {108}},\ \bibinfo {pages} {E67} (\bibinfo {year} {2011})}\BibitemShut
  {NoStop}%
\bibitem [{\citenamefont {Reder}(1988)}]{Reder1988}%
  \BibitemOpen
  \bibfield  {author} {\bibinfo {author} {\bibfnamefont {C.}~\bibnamefont
  {Reder}},\ }\href@noop {} {\bibfield  {journal} {\bibinfo  {journal} {J.
  Theor. Biol.}\ }\textbf {\bibinfo {volume} {135}},\ \bibinfo {pages} {175}
  (\bibinfo {year} {1988})}\BibitemShut {NoStop}%
\bibitem [{\citenamefont {Sauro}\ and\ \citenamefont
  {Ingalls}(2004)}]{Sauro2004}%
  \BibitemOpen
  \bibfield  {author} {\bibinfo {author} {\bibfnamefont {H.~M.}\ \bibnamefont
  {Sauro}}\ and\ \bibinfo {author} {\bibfnamefont {B.}~\bibnamefont
  {Ingalls}},\ }\href@noop {} {\bibfield  {journal} {\bibinfo  {journal}
  {Biophys. Chem.}\ }\textbf {\bibinfo {volume} {109}},\ \bibinfo {pages} {1}
  (\bibinfo {year} {2004})}\BibitemShut {NoStop}%
\bibitem [{\citenamefont {Sauro}\ and\ \citenamefont {Fell}(2000)}]{Sauro2000}%
  \BibitemOpen
  \bibfield  {author} {\bibinfo {author} {\bibfnamefont {H.~M.}\ \bibnamefont
  {Sauro}}\ and\ \bibinfo {author} {\bibfnamefont {D.~A.}\ \bibnamefont
  {Fell}},\ }in\ \href@noop {} {\emph {\bibinfo {booktitle} {Animating the
  Cellular Map: Proceedings of the 9th International Meeting on
  BioThermoKinetics}}}\ (\bibinfo {organization} {Stellenbosch University
  Press},\ \bibinfo {year} {2000})\ pp.\ \bibinfo {pages}
  {221--228}\BibitemShut {NoStop}%
\bibitem [{\citenamefont {{R Development Core Team}}(2008)}]{R}%
  \BibitemOpen
  \bibfield  {author} {\bibinfo {author} {\bibnamefont {{R Development Core
  Team}}},\ }\href@noop {} {\emph {\bibinfo {title} {{R: A language and
  environment for statistical computing}}}}\ (\bibinfo  {publisher} {R
  Foundation for Statistical Computing},\ \bibinfo {address} {Vienna,
  Austria},\ \bibinfo {year} {2008})\BibitemShut {NoStop}%
\bibitem	[]{sbolv}%
  \BibitemOpen
  \bibfield  {author} {\bibinfo {author} {\bibnamefont {{J. Quinn, 	J. Beal, S. Bhatia, P. Cai, J. Chen, K. Clancy, N. Hillson, M. Galdzicki, A. Maheshwari, P Umesh, M. Pocock, C. Rodriguez, G.-B. Stan, D. Endy}}},\ }\href@noop {} {\bibfield  {journal} {\bibinfo  {journal}
  {BioBricks Foundation RFP 93}\ }
  (\bibinfo {year} {2013})}\BibitemShut {NoStop}%
\bibitem []{SI}%
  \BibitemOpen
	  {\bibinfo {booktitle} {See Supplementary Material Document No.xxxxxx  for more detailed discussions on the derivation of  Eqs.~\eqref{eqn:vtilde}  and \eqref{variance2} and other numerical results. For information on Supplementary Material, see http://www.aip.org/pubservs/epaps.html}}\BibitemShut {NoStop}%
\end{thebibliography}

%

\pagebreak

\pagebreak

\pagebreak

\pagebreak

\pagebreak

\pagebreak

\end{document}